\documentclass[
  journal=pasa,
  manuscript=research-paper, 
  year=2020,
  volume=37,
]{cup-journal}

\usepackage{microtype,siunitx,booktabs}
\definecolor{structure@color}{RGB}{0,32,135}
\usepackage{hyperref}
\hypersetup{colorlinks=true,linkcolor=structure@color,citecolor=structure@color,filecolor=structure@color,urlcolor=structure@color}
\title{The role of impact parameter in typical close galaxy flybys}

\author{A. Mitra{\v s}inovi{\' c}}
\affiliation{Astronomical Observatory, Volgina 7, 11060 Belgrade, Serbia}
\alsoaffiliation{Department of Astronomy, Faculty of Mathematics,
University of Belgrade, Studentski trg 16, 11000 Belgrade,
Serbia}
\email[A. Mitra{\v s}inovi{\' c}]{amitrasinovic@aob.rs}

\author{M. Micic}
\affiliation{Astronomical Observatory, Volgina 7, 11060 Belgrade, Serbia}

\doi{10.1017/pasa.2020.32}

\received {dd Mmm YYYY}
\revised  {dd Mmm YYYY}
\accepted {dd Mmm YYYY}
\published{22 September 2020}


\begin{document}

\begin{abstract}
Close galaxy flybys, interactions during which two galaxies inter-penetrate, are frequent and can significantly affect the evolution of individual galaxies. Equal-mass flybys are extremely rare and almost exclusively distant, while frequent flybys have mass ratios $q=0.1$ or lower, with a secondary galaxy penetrating deep into the primary. This can result in comparable strengths of interaction between the two classes of flybys and lead to essentially the same effects. To demonstrate this, emphasise and explore the role of the impact parameter further, we performed a series of $N$-body simulations of typical flybys with varying relative impact parameters $b/R_{\mathrm{vir},1}$ ranging from $0.114$ to $0.272$ of the virial radius of the primary galaxy. Two-armed spirals form during flybys, with radii of origin correlated with the impact parameter and strengths well approximated with an inverted S-curve. The impact parameter does not affect the shape of induced spirals, and the lifetimes of a distinguished spiral structure appear to be constant, $T_\mathrm{LF} \sim 2$ Gyr. Bars, with strengths anti-correlated with the impact parameter, form after the encounter is over in simulations with $b/R_{\mathrm{vir},1} \leq 0.178$ and interaction strengths $S\geq0.076$, but they are short-lived except for the stronger interactions with $S\geq0.129$. We showcase an occurrence of multiple structures (ring-like, double bar) that survives for an exceptionally long time in one of the simulations. Effects on the pre-existing bar instability, that develops much later, are diverse: from an acceleration of bar formation, little to no effect, to even bar suppression. There is no uniform correlation between these effects and the impact parameter, as they are secondary effects, happening later in a post-flyby stage. Classical bulges are resilient to flyby interactions, while dark matter halos can significantly spin up in the amount anti-correlated with the impact parameter. There is an offset angle between the angular momentum vector of the dark matter halo and that of a disc, and it correlates linearly with the impact parameter. Thus, flybys remain an important pathway for structural evolution within galaxies in the local Universe.
\end{abstract} 

\section{INTRODUCTION}
\label{sec:int}
In $\Lambda$CDM universe, mergers are the driving formation and evolution mechanism of galaxies. As a result, interactions that do not end in a merger were explored to a lesser extent. That does not imply non-merger interactions are of less importance. By now, it has become common knowledge that interactions and tidal effects can produce various tidal and morphological structures \citep{tt1972,Eneev1973,barnes&hernquist1992,Tutukov2006,Dubinski&Chakrabarty2009} we regularly observe in galaxies. Clusters of galaxies, given their nature, are ideal environments for frequent interactions, especially penetrating encounters between satellites \citep{tormen1998,Knebe2004PASA}. In addition to the collective effects of the cluster itself, \cite{gnedin2003} found that peaks of the tidal force do not always correspond to the closest approach to the cluster centre, but instead to the local density structures (e.g. massive galaxies or the unvirialised remnants of in-falling groups of galaxies). These galaxy-galaxy interactions have comparable effects on the evolution of late-type galaxies as galaxy-cluster interaction \citep{hwang2018}. Additionally, multiple high-speed encounters between galaxies within clusters, the process known as galaxy harassment \citep{moore1996}, can cause morphological transformations with galaxies moving through the Hubble sequence from late-type discs to dwarf spheroidals \citep{moore1998,moore1999,mastropietro2005}. It is, thus, clear that the role of non-merger interactions in the evolutionary process of galaxies should not be underestimated.

However, there is no rigid classification of interactions, as almost any non-merger could be classified as a flyby. Not so close, distant encounters are naturally weaker but happen more frequently. For their impact on the evolution of galaxies to be substantial, usually, multiple such events have to occur \citep{hwang2018}. Conversely, despite being less frequent, closer penetrating encounters have the potential to leave a lasting imprint on interacting galaxies in a single event. Several studies examined rates and types of interactions in a cosmological framework \citep{sinha2012,lhuillier2015,shan2019}, leading to similar conclusions. While the number of close flybys is comparable or lower than the number of mergers at higher redshifts, towards lower redshifts (e.g. $z \leq 2$) these flybys outnumber mergers, particularly for less massive haloes and in high-density environments. In their follow-up study \citet{sinha2015} further explored interaction parameters defining typical flybys. They found that, in the majority of flybys secondary, intruder halo penetrates deeper than the half-mass radius of the primary $\sim R_\mathrm{half}$ with initial relative velocity $\sim 1.6 \times V_\mathrm{vir}$ of the primary and the typical mass ratio of interacting haloes is $q\sim 0.1$.

Given that frequency and strength of close galaxy flybys suggest that these interactions have the potential to significantly affect the evolution of individual galaxies, with the largest contribution at the present epoch, more studies exploring the role of close flybys followed. \citet{kim2014} showed that flybys can create warps in the primary galaxy disc. Naturally, close flybys are not the only warp formation mechanism, as distant flybys or even gas accreted from infalling satellites can also contribute \citep[e.g.][]{gomez2017,semczuk2020}. Tidal encounters are, however, the most common mechanism for the formation of strong warps, and most warped galaxies reside in dense environments \citep{ann2016,reshetnikov2016}. In addition to warps, tidal perturbation induced by flybys can create ring structures \citep{Younger2008}, kinematically decoupled cores \citep{Hau1994MNRAS,DeRijcke2004} and change the direction of galaxy's angular momentum \citep{bett2012,Cen2014spin,Lee2018ApJspin}. Secondary galaxies can even, through the process of tidal stripping in extreme cases of these interactions, become dark matter-deficient \citep[e.g.][]{ogiya2018,shin2020,jackson2021,maccio2021,moreno2022}.

It is well known that one formation channel of galactic bars involves interactions between galaxies \citep{Noguchi1987,Miwa1998,Berentzen2004}, and several authors investigated bar formation in close galaxy flybys. \citet{lang2014} found that bar forms in both galaxies in flybys with mass ratio $q=1$, and only in secondary with mass ratio $q=0.1$, and noted that induced changes are significantly stronger in flybys with prograde orbits (as opposed to retrograde orbits). By contrast, \citet{martinezvalpuesta2017} reported that bars of similar strengths and sizes formed in both prograde and retrograde encounters. Their result is particularly unusual as it is long known \citep[ever since the work of][]{tt1972} that prograde encounters have a much stronger effect on the galactic structure than retrograde ones. \citet{lokas2018} resolved this controversy confirming the results of \citet{lang2014}, and demonstrated that the simple impulse approximation \citep[used by][]{martinezvalpuesta2017} is not a viable approach for studying the effects of galaxy flybys. Both \citet{lang2014} and \citet{lokas2018} reported that flybys also lead to the formation of two-armed spiral structure, in addition to bars. \citet{pettitt2018} carried out a detailed study of bars and spirals in galaxy flybys, utilising hydrodynamical simulations and using diverse galaxy models to achieve different interaction strengths in flyby simulations while keeping an impact parameter fixed. They found that these interactions can induce a wide variety of morphological features and concluded, among other things, that flybys might be responsible for more of the observed morphologies than previously expected. While also tackling spiral and bar formation, \citet{kumar2021} drew attention to the evolution of bulges, both non-rotating classical ones and pseudobulges. They reported that strong spiral arms form in all simulations, the disc thickens, and pseudobulges become dynamically hotter. However, the classical bulges in their simulations mostly remain unaffected - these structures are, thus, quite resilient to flyby interactions.

As discussed by \citet{oh2015}, (tidal) strength of interaction can be quantified either simply, taking into account only the mass ratio of interacting galaxies, the size of the perturbed galaxy and impact parameter $b$, or in a more complex form, taking into account the interaction timescale with \citet{elmegreen1991} parameter. Regardless of the chosen measure, the strength of interaction is the most sensitive to the impact parameter $b$ since it scales with it like $b^{-3}$, while the scaling is linear with the other parameters. Therefore, even subtle changes of impact parameter in typical, close galaxy flybys can lead to vastly different outcomes. While \citet{kumar2021} addressed different pericentres (i.e. impact parameters), they did so using three different values ranging from $40$ kpc to $80$ kpc. To explore the role of impact parameter in typical close galaxy flybys further, we plan to sample more values ranging from the lowest one used in similar research up to the half-mass radius of the primary galaxy. This way, we can also attempt to determine the functional dependence of various effects on impact parameter.

The paper is organised as follows. In Section~\ref{sec:models} we define our models and simulations and discuss initial conditions. In Section~\ref{sec:qualdisc} we perform a qualitative analysis of the primary galaxy's disc structure, while in subsequent sections we focus closely on quantifying and discussing specific structures: spiral arms in Section~\ref{sec:spirals}, and bars in Section~\ref{sec:bars}. We tackle the effects on spherical components of the galaxy, dark matter halo and stellar bulge, in Section~\ref{sec:bulgehalo}. Finally, we discuss possible implications in Section~\ref{sec:impdis} and summarise our results and conclusions in Section~\ref{sec:sum}.

\section{MODELS AND SIMULATIONS}
\label{sec:models}
We used these simulations to examine the intruder's dark matter mass loss \citep{mitrasinovic2022}. In this paper, our focus is on the primary galaxy and its structural evolution. Thus, we will overview our models and simulations, emphasising parts relevant to this work. We used \texttt{GalactICs} software package \citep{kuijken1995,widrow2005,widrow2008} for constructing our galaxy initial conditions. In general, the code generates a self-consistent galaxy model which can consist of up to three components: NFW \citep{nfw1997} dark matter halo, exponential stellar disc, and \citet{hernquist1990} stellar bulge.

Our primary galaxy (which will be referred to as, simply, galaxy) consists of all three components. Dark matter halo with $N_\mathrm{H} = 6\cdot 10^5$ particles has total mass $M_\mathrm{H} = 9.057 \cdot 10^{11} \mathrm{M}_\odot$, scale length $a_\mathrm{H} = 13.16$ kpc, and concentration parameter $c = 15$. Stellar disc with $N_\mathrm{D} = 3\cdot 10^5$ particles has total mass $M_\mathrm{D} = 7.604 \cdot 10^{10} \mathrm{M}_\odot$, scale length $R_\mathrm{D} = 5.98$ kpc, scale height $z_\mathrm{D} = 0.688$ kpc, and central velocity dispersion $\sigma_{R_0} = 98.9$ km s$^{-1}$. \citet{toomre1964} parameter of our disc is $Q (2.5 \cdot R_\mathrm{D}) = 1.73$, which should make the disc fairly stable against bar formation at least for a few Gyr. To additionally stabilise the disc against bar formation, we include a rather massive bulge component \citep{shen2004, athanassoula2005}: stellar bulge with $N_\mathrm{B} = 1\cdot 10^5$ particles has total mass $M_\mathrm{B} = 2.502 \cdot 10^{10} \mathrm{M}_\odot$, and scale radius $R_\mathrm{B} = 2.182$ kpc. This model is somewhat heavy compared to typical Milky Way models \citep[see][and references therein]{wang2020} with the larger and more massive stellar disc and massive stellar bulge, more akin to those of the Andromeda galaxy \citep[e.g.][]{kafle2018}. The choice of parameters is suitable for this study, as we are considering a general massive disc galaxy experiencing a flyby encounter.

In this galaxy model, dark matter halo particles are heavier than those of stellar components (disc and bulge), with particle mass ratio $m_\mathrm{H}/m_\mathrm{DB} \simeq 6$, where $m_\mathrm{H}= 15 \cdot 10^5$ M$_\odot$ is dark matter halo particle mass, and $m_\mathrm{DB} \simeq 2.5 \cdot 10^5$ M$_\odot$ is particle mass of stellar components. The value of particle mass ratio is well within the range of values ($1.6$,$22.2$) used in similar, relevant research \citep[e.g.][]{lang2014,pettitt2018}. Additionally, \citet{oh2015} tested the dependence of simulation outcomes on the particle resolution, comparing the low-resolution model of $N_\mathrm{tot} = 5.1 \cdot 10^5$ with the high-resolution model of $N_\mathrm{tot} = 10.2 \cdot 10^6$ total particles, and found that the resolution does not make a significant difference in the properties of tidally-induced features.

For the sake of simplicity, our secondary galaxy (which will be referred to as intruder) consists of dark matter halo, and stellar bulge. The intruder model was scaled to be $10$ times smaller than the galaxy model, in both the number of particles and total mass. This results in dark matter halo with $N_\mathrm{H} = 6\cdot 10^4$ particles of total mass $M_\mathrm{H} = 9.044 \cdot 10^{10} \mathrm{M}_\odot$, scale length $a_\mathrm{H} = 4.578$ kpc, and concentration parameter $c = 20$. Stellar bulge with $N_\mathrm{B} = 4\cdot 10^4$ particles has total mass $M_\mathrm{B} = 1.022 \cdot 10^{10} \mathrm{M}_\odot$, and scale radius $R_\mathrm{B} = 3.145$ kpc.

Both models were evolved for 5 Gyr in isolation using publicly available code \texttt{GADGET2} \citep{springel2000,springel2005gadget}, compiled with the option to calculate and output particle potential energy, with outputs recorded every 0.01 Gyr. The purpose of this is twofold. Firstly, we need to ensure the models are stable enough. To consider the model stable, we require that energy and angular momentum are conserved, and mass distributions of subsystems do not change significantly. And secondly, we will compare the results of flyby simulations with our galaxy in isolation to differentiate flyby-induced features or changes from those arising from secular evolution. General stability requirements are met for both models: energy and angular momentum change for less than $1\%$ throughout the simulation, and mass distributions remain the same. However, the basic assessment of mass distributions, through comparing mass and density radial profiles, does not provide enough information on bar formation in galaxy disc. To account for this, we employ the same method which we use for bar detection in flyby simulations, described in Section~\ref{sec:bars}. A weak bar starts to show after 3 Gyr but only grows, slowly and steadily, after 4 Gyr until the end of the simulation when its strength is still low. This formation time is long after the interaction occurs in flyby simulations (Subsection~\ref{sec:flybysim}), so we can consider the galaxy stable against bar formation. However, aside from studying the possible early bar formation as a direct consequence of close galaxy flybys, we can also examine the effect of flybys on this pre-existing bar instability.

For all our simulations, including flyby ones, we use fixed gravitational softening length parameter, $\epsilon = 0.05$ kpc for all particle types. In general, value of the softening length parameter scales with the number of particles $N$ and dimension of the system $R$ as $R/N^{1/2} < \epsilon < R/N^{1/3}$ \citep{b&t2008}. However, in practice, the optimal value for softening length parameter remains somewhat ambiguous as several criteria have been proposed \citep[e.g.][]{merritt1996,dehnen2001,power2003,zhang2019}. The lowest optimal value of the softening length parameter yields $\epsilon_{\mathrm{dm}} \approx 0.25$ kpc for dark matter particles, and $\epsilon_{\mathrm{b}} \approx 0.05$ kpc for baryon particles in our models. We performed test simulations with these values, and a fixed value of $\epsilon = 0.05$ kpc for all particle types for the galaxy model and found no significant differences in disc evolution and bar formation over 5 Gyr. Computational time is, however, significantly lower for simulation with fixed $\epsilon$. \cite{iannuzzi2013} also find that adopting the fixed softening length does not affect the evolution of the inner disc component. 

\subsection{FLYBY SIMULATIONS}
\label{sec:flybysim}
To reach the pericentre distance as soon as possible and follow the evolution of galaxy disc long after the encounter, galaxy and intruder are initially set as a contact system. Distance from their centres is equal to the sum of their virial radii, $d = R_{\mathrm{vir},1} + R_{\mathrm{vir},2} \approx 290$ kpc. Galaxy remains static in the centre of the simulation box, while the intruder is set on a prograde orbit with initial relative velocity $v_{0} = 500$ km s$^{-1}$, co-planar with the galaxy disc. By slightly varying angles of initial position and intruder velocity vector, we achieved different pericentres (impact parameters) $b$. We define duration of the interaction as time during which galaxy and intruder overlap i.e. distance between their centres is $d \leq R_{\mathrm{vir},1} + R_{\mathrm{vir},2}$. Duration of the interaction remains the same in all simulations, $1.08$ Gyr, and pericentre occurs at the same time ($0.56$ Gyr) during simulations. Pericentre distances (impact parameters) $b$ and velocities $v_b$ are different.

In addition to pericentre distance, we use the Elmegreen tidal strength parameter $S$ \citep{elmegreen1991} to characterise the strength of the interaction:
\begin{equation}\label{eq:elmegreen_s}
S = \frac{M_\mathrm{int}}{M_\mathrm{gal} \big(R<R_{\mathrm{gal}}\big)} \Bigg( \frac{R_\mathrm{gal}}{b} \Bigg)^{3} \frac{\Delta T}{T}
\end{equation}
where $M_\mathrm{int}$ is intruder (secondary galaxy) mass, $b$ pericentre, $R_\mathrm{gal}$ represents disc truncation radius (in our case $R_\mathrm{gal}=32$ kpc), and $M_\mathrm{gal} \big(R<R_{\mathrm{gal}}\big)$ is total galaxy mass inside $R_\mathrm{gal}$. $\Delta T$ is the time it takes the intruder to travel one radian around the galaxy centre near the pericentre, and $T$ the time it takes stars in the outer parts of the stellar disc, on  $R = R_\mathrm{gal}$, to travel one radian around the centre and can be calculated as $T = (R_\mathrm{gal}^3/GM_\mathrm{gal})^{1/2}$. The range of $S$ we cover is comparable to the range $0.01<S<0.25$ several studies found to be the optimal for producing any sort of spiral arms \citep{elmegreen1991,oh2008,oh2015,Pettitt2016,semczuk2017}.

Relevant interaction parameters are listed in Table~\ref{tab:flyby_table}. Simulations were named after our rough estimate of the impact parameter, which evidently differs from the actual impact parameter (column $b$). These simulations cover deep flybys, with impact parameters ranging from $0.114 \cdot R_{\mathrm{vir},1}$ to slightly over galaxy's half-mass radius ($\sim 49$ kpc). Impact parameter range, combined with the mass ratio of $10:1$ (i.e. $q=0.1$), is in line with what \citet{sinha2015} found to be the typical interaction parameters of galaxy flybys, while our adopted initial relative velocity is higher than the most common one. \cite{gnedin2003} reported that, in Virgo-type cluster simulation, relative velocities of interacting galaxies show a skewed distribution, peaking at $\sim 350$ km s$^{-1}$, with median $\sim 800$ km s$^{-1}$ and mean value $\sim 1000$ km s$^{-1}$. This was used by \citet{kim2014} as a basis for their choice of initial relative velocity, $v_0 = 600$ km s$^{-1}$. Thus, value of $v_0 = 500$ km s$^{-1}$ adopted here, despite not being the most common one, can still be considered as representative and realistic.

\begin{table}
	\centering
	\caption{List of flyby simulations with their pericentre distances (impact parameters) $b$ and velocities $v_b$, impact parameters $b/R_{\mathrm{vir},1}$ relative to the virial radius of the primary galaxy $R_{\mathrm{vir},1}$ and the Elmegreen parameter $S$ defined with equation~\ref{eq:elmegreen_s}.}
	\label{tab:flyby_table}
	\begin{tabular}{lcccc}
		\hline
		Name & $b$ [kpc] & $b/R_{\mathrm{vir},1}$ & $v_b$ [km s$^{-1}$] & $S$\\
		\hline
		B30 & 22.50 & 0.114 & 660.14 & 0.177 \\
		B35 & 26.53 & 0.135 & 650.86 & 0.129 \\
		B40 & 30.69 & 0.156 & 641.80 & 0.098 \\
		B45 & 35.07 & 0.178 & 632.86 & 0.076 \\
		B50 & 39.62 & 0.201 & 624.25 & 0.060 \\
		B55 & 44.27 & 0.224 & 616.16 & 0.049 \\
		B60 & 48.99 & 0.248 & 608.09 & 0.040 \\
		B65 & 53.72 & 0.272 & 601.28 & 0.034 \\
		\hline
	\end{tabular}
\end{table}



\section{QUALITATIVE ANALYSIS OF THE DISC}\label{sec:qualdisc}
\begin{figure*}
\includegraphics[width=0.99\textwidth,keepaspectratio=true]{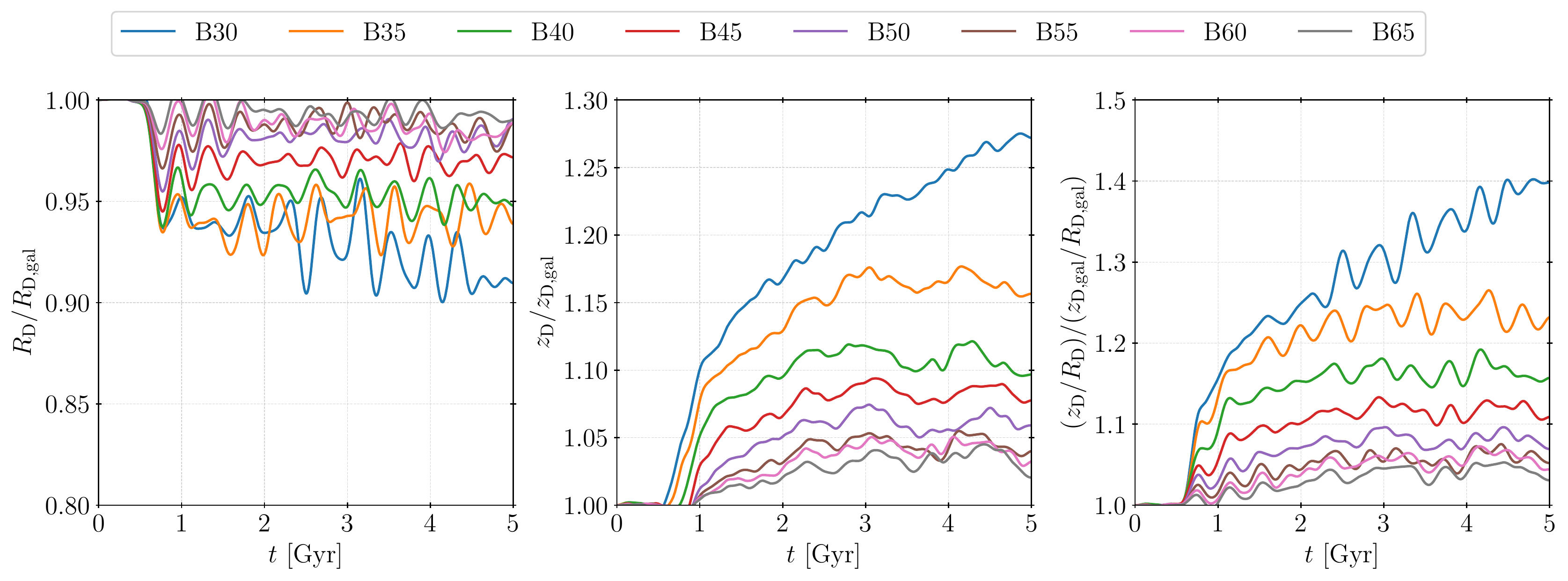}
\caption{Evolution of the disc scale length $R_\mathrm{D}$ (left panel), scale height $z_\mathrm{D}$ (middle panel) and thickness $z_\mathrm{D}/R_\mathrm{D}$ (right panel) in flyby simulations relative to the isolated case. Different simulations are represented with different line colours.}
\label{discscale}
\end{figure*}
Our data pre-processing starts with centring galaxy disc on a particle with the lowest potential energy, rather than its centre of mass, to avoid errors in centre of mass determination caused by particles that left the disc or elongated tidal tails. This step is crucial in simulations with stronger interactions: the difference between the location of the lowest potential energy particle and simple centre of mass determination can be significant, reaching 2-3 kpc. Additionally, we rotate the galaxy aligning the angular momentum vector of the disc with a positive direction of the $z$-axis, ensuring the disc plane lies on the $x-y$ plane. We exclude particles farther above and below the disc plane ($|z|> 3$ kpc) from the further analysis. For each simulation output (snapshot), we calculate the disc 2D density profile and use exponential fit to determine the value of disc scale length $R_\mathrm{D}$. We also calculate the disc vertical density profile in the inner region (where $R<10$ kpc) and fit the $\mathrm{sech}^{2}$ function to determine the value of disc scale height $z_\mathrm{D}$.

The evolution of these disc scales, as well as the disc thickness $z_\mathrm{D}/R_\mathrm{D}$ in flyby simulations is shown in Figure~\ref{discscale} relative to their counterpart in isolation. The decrease of scale length and increase of scale height is noticeable immediately after the intruder reaches the pericentre. Both of these effects contribute to disc thickening. The change of both scales anti-correlates with the impact parameter: it is more pronounced in closer flybys. In the simulation with the closest flyby, B30, the change of scale length is highly variable, decreasing up to $\sim 10\%$ compared to the isolated case, and the scale height continues to grow up to $\sim 27\%$ at the end of the simulation, which results in $\sim 40\%$ change in disc thickness. These values hint that the bar forms in B30 and continuously evolves, growing in length and strength and buckling up.

While the evolution of scale length we observe is comparable to the results of \citet{kumar2021}, the evolution of scale height is not. They note that disc thickening is a result of the decrease in scale length. We find the opposite: the increase in disc height is larger and contributes more to the disc thickening. The reason for this difference is due to the fact that their model with the classical bulge is bar-stable, and the changes in disc scales are mainly driven by the formation and evolution of spiral arms. In our simulations, bars most likely form and evolve along with the spiral arms and significantly affect the disc height in the inner disc region.

\subsection{NON-AXISYMMETRIC STRUCTURES}
\begin{figure*}
\includegraphics[width=0.82\textwidth,keepaspectratio=true]{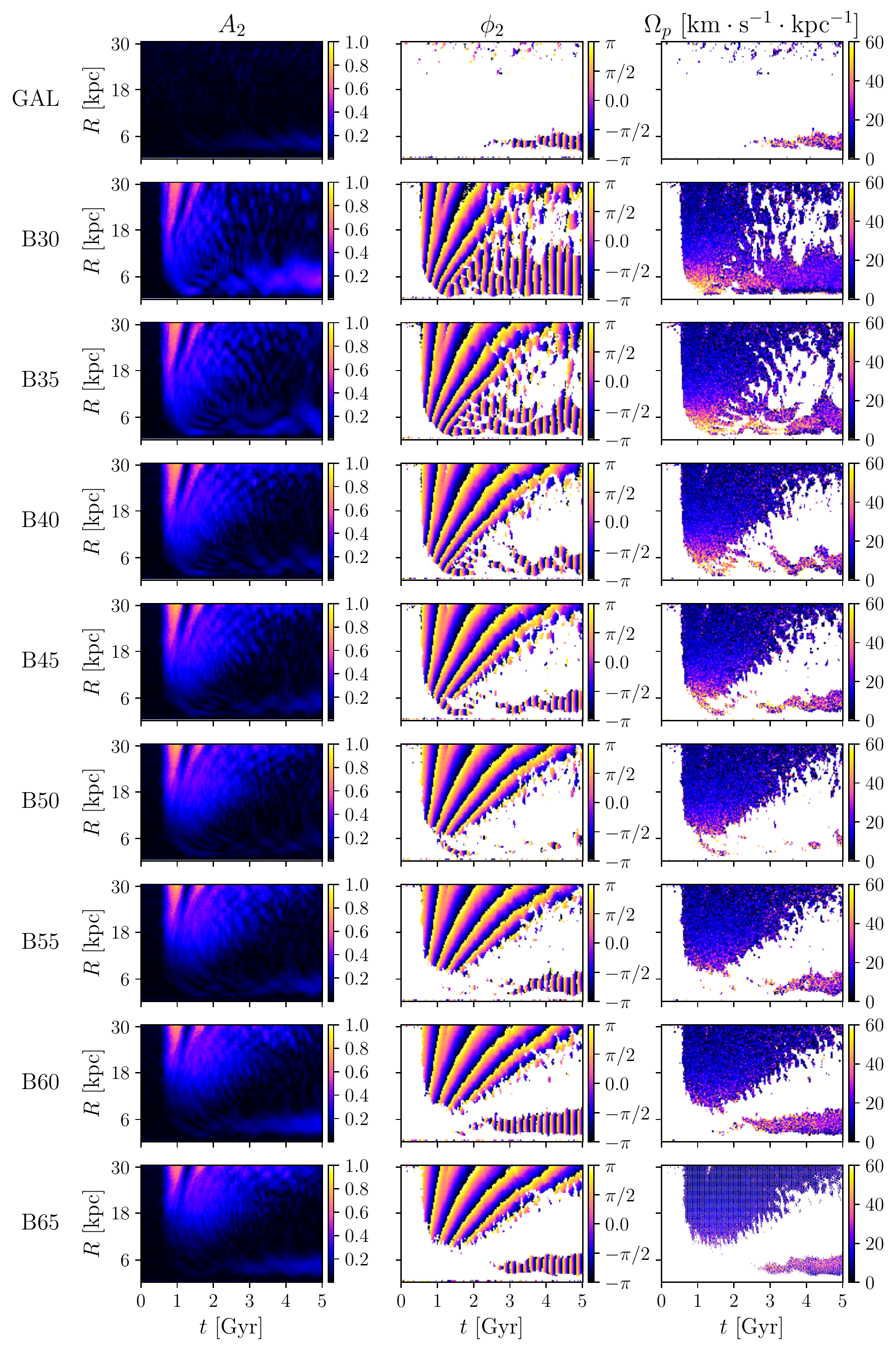}
\caption{Evolution maps of Fourier second mode: amplitude $A_2 (t,R)$ (left column panels), phase $\phi_2 (t,R)$ (middle column panels) and pattern frequency $\Omega_p (t,R)$ (right column panels). GAL row shows isolated galaxy simulation, while the rest are showing flyby simulations, as defined in Table~\ref{tab:flyby_table}. We mask regions where $A_2<0.1$ on phase angle and pattern frequency maps to eliminate the noise.}
\label{fmaps}
\end{figure*}
In essence, our method is based on Fourier analysis commonly used for detection of morphological structures \citep[e.g.][]{athanassoula2002,athanassoula2013}. When disc is decomposed in Fourier modes, prominent structures like bars and two-armed spirals contribute the most to the second mode $m=2$. Relative (mass normalised, since zeroth mode $C_0$ yields total mass) second mode $C_2$ is calculated as:
\begin{equation}\label{eq:rel_c2}
\frac{C_2}{C_0}=\frac{1}{M} \sum^N_{j=1} m_j e^{2i\phi_j} = C_{21} + i C_{22}
\end{equation}
\noindent where $M$ is the total mass and summation is performed over all particles with masses $m_j$ and angles $\phi_j$ (cylindrical coordinate) in $x-y$ plane. With $C_{21}$ and $C_{22}$ representing real and imaginary part respectively, of its complex form, amplitude $A_2$ and phase $\phi_2$ are then calculated as:
\begin{equation}\label{eq:a2_p2}
A_2 = \sqrt{C_{21}^2+C_{22}^2}
\quad\text{and}\quad 
\phi_2 = \mathrm{arctan} \bigg(\frac{C_{22}}{C_{21}}\bigg)
\end{equation}
\noindent This calculation can be performed globally (accounting for all disc particles) or semi-globally (accounting for particles residing in a specific broad radial region). In that case, high amplitude $A_2$ would indicate that some kind of non-axisymmetric morphological feature or more of them are present, while phase angle $\phi_2$ would be of little use, giving only a rough estimate of the position angle for the most prominent feature or its part.

More detailed information about the feature type (whether it's a bar or two-armed spirals) and its region, can be obtained by performing this calculation locally, slicing the disc in annuli in the $x-y$ plane for each simulation snapshot. This would result in evolution maps for both amplitude $A_2 (t,R)$, and phase $\phi_2 (t,R)$ where $R$ is the central radius of each annulus and $t$ time of each snapshot. Upon first inspection, these evolution maps can provide valuable insight into structure formation and evolution. A high enough amplitude $A_2$ is a characteristic of both bars and two-armed spirals. The behaviour of phase angle $\phi_2$ along the radial axis helps differentiate between the two. Bars have an almost constant phase angle, while the phase angle of spiral arms is uniformly changing. Additionally, we calculate pattern frequency evolution maps $\Omega_p (t,R)$ based on phase angle temporal changes, and convert them to appropriate units, $[\mathrm{km}$ $\mathrm{s^{-1}}$ $\mathrm{kpc}^{-1}]$.

We utilise these evolution maps as a part of our initial qualitative analysis and employ additional methods for the quantitative analysis described in the following appropriate Sections. Evolution maps are shown in Figure~\ref{fmaps} with columns (left to right) for amplitude $A_2 (t,R)$, phase angle $\phi_2 (t,R)$ and pattern frequency $\Omega_p (t,R)$. Different rows correspond to different simulations defined in Table~\ref{tab:flyby_table} while the first row, denoted as GAL, represents the isolated simulation of our galaxy model. To eliminate the noise on phase angle and pattern frequency evolution maps, we mask regions where the amplitude is low ($A_2<0.1$). In all flyby simulations, tidal tails and remnants of the induced spiral arms (spiral arcs) are visible in the outer parts of the disc (on $R\geq 24$ kpc) until the end as amplitude is still moderately high, and phase angle changes uniformly. The higher noise in these regions in simulations with stronger, closer flybys (compared to simulations with weaker, farther flybys) suggests that remnants of spiral arms are slightly more stable as the impact parameter increases.

The first peak in amplitude (when the intruder reaches its pericentre) signifies the formation of the elongated tidal tail. Shortly after that and before the encounter is even over two-armed spiral structure forms. There seems to be a clear correlation (and anti-correlation) between the impact parameter and the strength and radius of origin of induced spirals. As the impact parameter increases, the strength of the spiral arms gets slightly weaker, and their origin shifts towards the outer disc parts. The impact parameter most likely does not affect the shape of spiral arms (pitch angle), as the phase profiles at formation time appear roughly the same in all simulations. However, stronger spirals in simulations with closer flybys, especially their inner segments, have higher pattern frequencies suggesting that they might wind up and dissolve faster.

Contrary to spiral arms, bar formation and evolution appears much more chaotic. As a direct consequence of the flyby, a short bar forms almost immediately after the encounter is over in closer flyby simulations. The strength of that bar decreases with the increase of the impact parameter. After a period of turbulent bar evolution, at around 3 Gyr in simulation B30, the bar steadily grows in strength and length. In simulation B35, there appears to be a long period of double-bar or ring-like feature, likely caused by winded up spirals around the already formed bar. Such a feature is possibly also present in simulation B40 (albeit for brief period), in which the bar evolution overall is turbulent. Simulation B50 is particularly interesting - it appears that pre-existing bar instability is constantly being suppressed, and the bar never fully forms. In all other simulations, bar evolution is stable and steady. Formation of this bar is mainly caused by the pre-existing instability, as it happens long after the encounter. While there are some differences compared to its counterpart in isolation, there seems to be no uniform correlation between bar parameters (e.g. length, strength or formation time) and impact parameter.

\section{TWO-ARMED SPIRAL STRUCTURE}\label{sec:spirals}
The shape of tidally induced spirals is usually almost perfectly logarithmic. This fact makes it convenient to use the widely accepted method \citep[e.g.][]{oh2015,semczuk2017,kumar2021} discussed by  \cite{sellwood1984} and \cite{sellwood1986} for calculating the spiral strength and pitch angle. The strength of the spiral arms is given by the formula:
\begin{equation}\label{eq:sp_a}
    A (m,p) = \frac{1}{N} \sum_{j=1}^N \mathrm{exp}[i (m \phi_j + p \mathrm{ln}R_j)]
\end{equation}
\noindent where $m$ relates to the number of arms (for our case with two-armed spirals $m=2$), $p$ is a free parameter that relates to the spiral pitch angle $\alpha$ and summation is performed over all particles with cylindrical coordinates $\{R_j,\phi_j\}$ within the examined annulus $(R_\mathrm{s,min},R_\mathrm{s,max})$. The annulus should ideally be chosen from the middle of the disc to avoid contamination of results caused by the bar in the inner region. We use ($12$ kpc, $24$  kpc) as it roughly corresponds to the middle disc region $(2 R_\mathrm{D},4 R_\mathrm{D})$ at the start. Following the procedure used by \cite{puerari2000}, we chose $(-50,+50)$ for the $p$ interval with the step size $dp=0.25$. After determining $p_\mathrm{max}$ which maximises $A(2,p)$ (Equation~\ref{eq:sp_a}), the spiral strength is defined as $| A(2,p_\mathrm{max})|$ while the pitch angle is calculated as $\alpha = \mathrm{arctan}(2/p_\mathrm{max})$.

Over time, tidally induced spirals rapidly wind up and eventually dissolve. It results in exponential decay segments, visible on the spiral strength and pitch angle temporal profiles, $A_2(t)$ and $\alpha (t)$. We will exploit this fact to estimate the spiral lifetimes based on the periods of those distinct exponential decay segments. Note that the lifetimes estimated this way should be considered as an upper limit as the pitch angle can drop below $\alpha \simeq 10^\circ - 15^\circ$, the values of most observed spirals \citep{gga1993,Ma2002A&A,b&t2008,mo&vdBosch&White2010}.
\subsection{RESULTS}
\begin{figure}
\includegraphics[width=0.95\columnwidth,keepaspectratio=true]{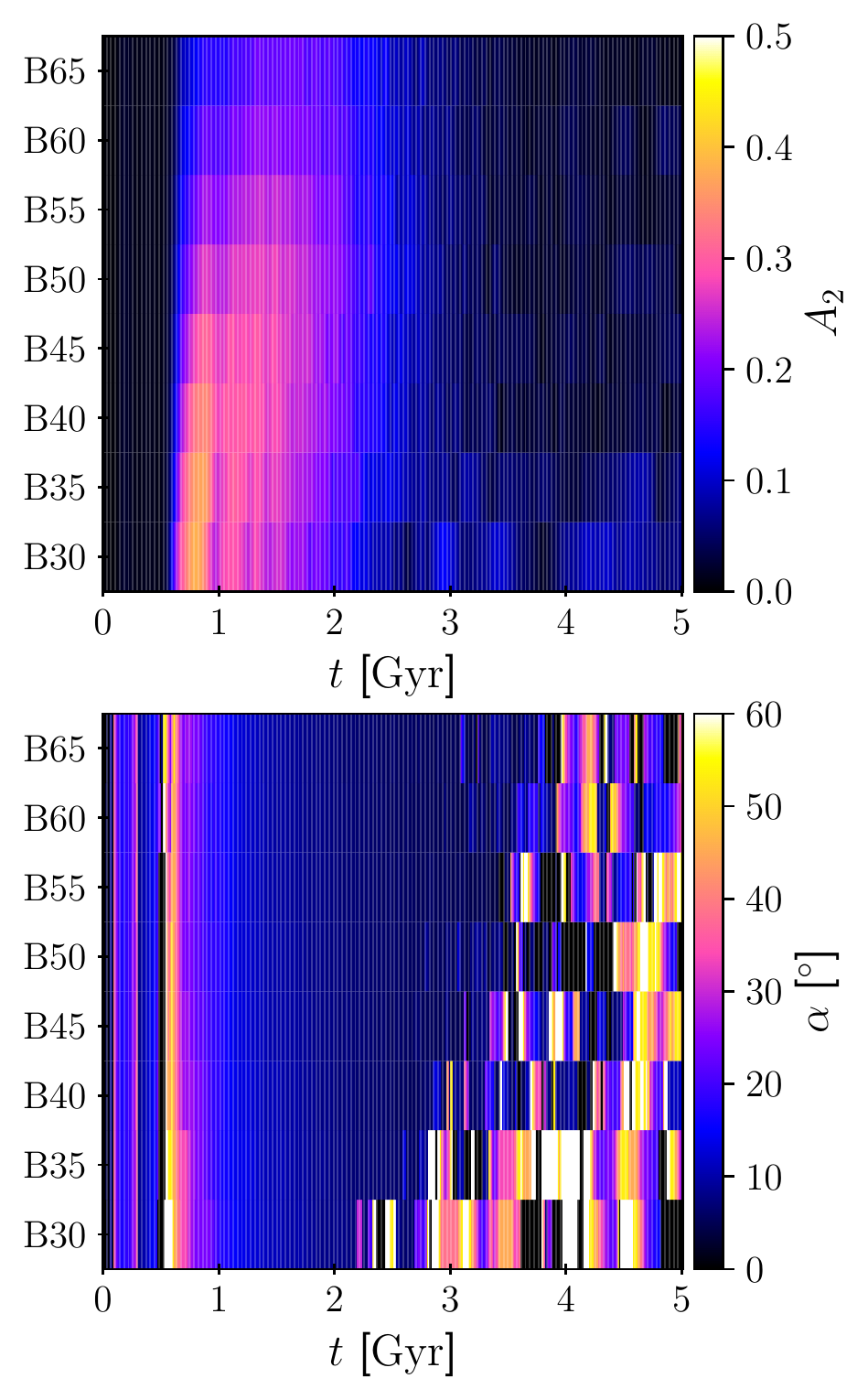}
\caption{Evolution of spiral parameters described in Section~\ref{sec:spirals}: the strength $A_2$ (upper panel) and the pitch angle $\alpha$ (lower panel).}
\label{pitch}
\end{figure}
\begin{figure}
\includegraphics[width=0.95\columnwidth,keepaspectratio=true]{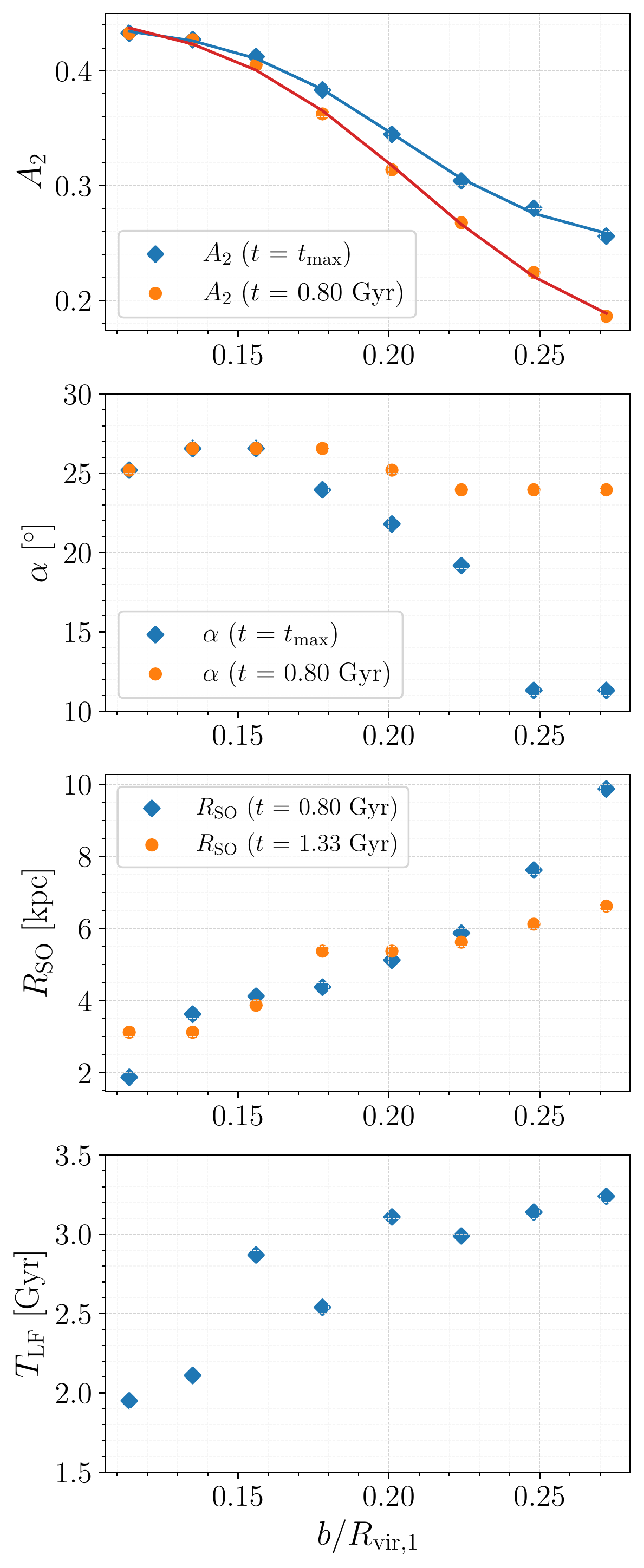}
\caption{Visual representation of spiral relevant parameters. Top to bottom: spiral strength $A_2$, pitch angle $\alpha$, spiral radius of origin $R_\mathrm{SO}$ and lifetime period $T_\mathrm{LF}$.}
\label{spiralparam}
\end{figure}
The evolution of spiral parameters described in this section is shown in Figure~\ref{pitch}, where $A_2$ is the spiral strength and $\alpha$ its pitch angle. Relevant parameters of the spirals are visually represented in Figure~\ref{spiralparam} as a function of impact parameter relative to the virial radius of the primary galaxy $b/R_{\mathrm{vir},1}$. These parameters include: maximum spiral strength $A_{2,\mathrm{max}}$ reached at time $t_\mathrm{max}$ (which varies between simulations) and with pitch angle $\alpha_\mathrm{max}$, spiral strength $A_2 (t)$ and pitch angle $\alpha (t)$ at roughly the formation time $t=0.8$ Gyr, all calculated based on data shown in Figure~\ref{pitch}. We estimated lifetime periods $T_\mathrm{LF}$ based on a behaviour of pitch angle temporal profiles $\alpha (t)$, which represent, as previously discussed, an upper limit for the lifetimes. Additionally, we estimated the radius of origin of the spirals $R_\mathrm{SO}$ as a minimum radius of distinct logarithmic segment of phase angle radial profiles $\phi_2 (R)$ (defined with Equation~\ref{eq:a2_p2}, included in Figure~\ref{fmaps} and discussed in previous Section) at two different times, $t = 0.8$ Gyr and $t = 1.33$ Gyr.  While distinguishable spirals might have formed slightly earlier, we adopt $t = 0.8$ Gyr as a formation time as it corresponds to the earliest maximum spiral strength.

Maximum spiral strengths $\{A_{2,\mathrm{max}}\}$ and strengths at roughly the formation time $\{A_2 (t=0.8$ Gyr$)\}$ have an evident pattern: they decrease with the increase of impact parameter. While the functional relationship is that of an inverted logistic S-curve in Figure~\ref{spiralparam}, cubic regression can be used alternatively. We exclude the parameters of those functions as they, most likely, depend on interaction parameters (e.g. mass ratio, initial relative velocity and implicitly interaction duration). Moreover, maximum spiral strengths are delayed more as the impact parameter increases. This is because the strongest tidal features form in the outer parts of the disc in weaker flybys, and they need time to evolve, wind up and migrate into the region where we calculate these parameters. The pitch angle panel in Figure~\ref{spiralparam} supports this: values at $t=0.8$ Gyr are comparable, while those at $t = t_\mathrm{max}$ are lower for higher impact parameters, implying that those spirals are more evolved. The evolution of the pitch angle $\alpha(t)$ in Figure~\ref{pitch} suggests that the impact parameter has little to no effect on the shape of the spirals, only on their strengths.

The spiral radius of origin $R_\mathrm{SO}$ is directly proportional to the impact parameter. However, the relationship is steeper at formation time $t=0.8$ Gyr, than at later time $t=1.33$ Gyr. This is due to two factors: in weaker, farther flybys, spiral arms evolve and wind up migrating into inner disc regions; in stronger, closer flybys, the spiral radius of origin shifts towards outer parts due to early bar formation. Note that these values should be considered as a lower limit since we approximated them based on the phase angle profiles disregarding the amplitude (i.e. the spiral strength). Should we include additional condition for the spiral strength, it would slightly shift $R_\mathrm{SO}$ towards higher values. Despite our crude approximation, it is undeniable that the spiral radius of origin is directly proportional to the impact parameter.

Lifetime periods $T_\mathrm{LF}$ are linearly increasing with the impact parameter, with two outliers, simulations B40 and B50. We previously mentioned that this estimate represents an upper limit, as the pitch angle can drop below the values of most observed two-armed spirals. Additionally, we should factor in the spiral strength $A_2$: it represents how much this feature is distinguishable. At later stages, $t\geq3$ Gyr, this value drops below $A_2 = 0.1$, and it is questionable whether such spirals could be easily resolved. More realistic lifetime periods (when the strength $A_2$ is sufficiently high) appear to be more or less constant in almost all simulations (B35-B65) $\sim 2$ Gyr, except for simulation B30 where it is shorter $\sim 1.5$ Gyr. Thus, the answer to does the impact parameter affect spiral lifetimes is complex: lifetimes of well resolved, prominent spirals, like the evolution of their shapes, seem to be independent of the impact parameter. However, those of any weak spiral feature or its remnant (especially in the outer disc part) have almost a linear correlation.
\subsection{DISCUSSION}
The results of our spiral arms analysis are mostly in line with the previous work on the subject. As \citet{pettitt2018} noted, stronger and thus closer (in our case) interactions tend to create more chaotic and disturbed discs. Two-armed spirals in those simulations are stronger and tend to dissolve faster, especially their inner parts. \citet{lokas2018} briefly discussed a similar observation: stronger and more persistent spirals were formed in weaker interactions. However, we warn that such a finding can be misleading and emphasise that this only applies to any spiral fragment or the remnant of spiral arms (i.e. if spiral arms are loosely defined). In case the spiral arms are more strictly defined, and we require that they are well-resolved, their lifetimes do not correlate with the impact parameter, remaining almost constant between simulations. That constant value should not be considered universal in galaxy flybys, as it possibly depends on other interaction parameters we kept fixed (e.g. interaction duration, intruder's orbital inclination).

Similarly, \citet{kumar2021} observed decay of the maximum spiral arms strength in their models with the classical bulge, which led to comparable strengths at the end in simulations with different impact parameters. The decaying nature of the spiral arms and their winding-up and dissolution, which we also observed, is not surprising. This scenario is, in fact, inevitable for tidally induced spirals \citep[e.g.][]{oh2008,oh2015}. Maintenance of well-resolved spiral arms typically requires repeated tidal perturbations caused by a satellite galaxy, multiple encounters or collective cluster effects. More specifically, while examining spiral arms in galaxies orbiting a Virgo-like cluster, \citet{semczuk2017} found that pericentre passages trigger the formation of spiral arms. These spirals wind up and dissipate only to be triggered again and regenerated during the next pericentre passage.

Bar formation and evolution (which we will discuss in the following Section) is considered as one of the spiral arms formation mechanisms \citep[e.g.][]{buta2009,salo2010}. While we did observe warping of the bar edges along its major axis at later stages (in three closest simulations: B30, B35 and B40), the effect is very brief: bars quickly lose their warps and later spiral formation does not happen. It seems to be in good agreement with \cite{diazgarcia2019}, for example, who found no strong observational evidence that spirals are bar-driven.

While remnants of the induced spiral structure and tidal tails are present in the disc outskirts until the end (in all our simulations), we did not detect any noticeable warp. At first, this might seem unusual, as close flybys are known to induce disc warps. However, both \citet{kim2014} and \citet{semczuk2020} found that most warps form in flybys with inclined orbits of the intruder galaxy. In planar encounters, like the ones we examined, disc warps are not expected as there is little to no tidal force acting in the vertical direction on the galaxy.

Idealised $N$-body simulations we performed are limited as they lack gas and its proper treatment. In the presence of gas, star-forming is concentrated in spiral arms \citep{kim2020} and even enhanced \citep{pettitt2017,yu2021}, which can additionally prolong the lifetimes of spirals and make spiral patterns more prominent and stronger. However, our estimated spiral lifetimes are not considerably short, even for well-resolved spiral arms. They are long enough for the intruder to leave the proximity of the primary galaxy. While the interaction-driven nature of grand design spirals is well known, there is still a number of grand design spiral galaxies lacking observable perturbers \citep{kendall2011,kendall2015}. We, thus, argue that very close flybys are ideal candidates for explaining such cases.

\section{BAR FORMATION AND EVOLUTION}\label{sec:bars}
The strength of the bar is frequently measured by the maximum value of amplitude $A_2$ (Equation~\ref{eq:a2_p2}), with a certain threshold. The threshold is not strictly defined: as the commonly used value of $A_2>0.2$ \citep[e.g.][]{aguerri2009,kazantzidis2011,gajda2018} can be biased towards strong bars, the value of $A_2>0.15$ is becoming adopted regularly \citep[e.g.][]{peschken2019,zhou2020}, while the usage of even lower values is not unheard of \citep{lang2014,lokas2018}. Instead of calculating the absolute maximum value of $A_2(R)$ profile, it is better to detect the local maximum in the inner parts of the disc (e.g. where $R<2 R_\mathrm{D}$) to avoid contamination caused by spiral arcs or tidal tails from the outer disc parts. This procedure is typically sufficient for the determination of the bar strength.However, the challenges of the bar length estimation persist in the presence of two-armed spirals.

\citet[][Section 8]{athanassoula2002} discussed different ways to estimate the bar length, with some of them based on Fourier decomposition. As amplitude profile $A_2(R)$ declines after reaching its maximum value $A_{2,\mathrm{max}}$, it is convenient to find the radius where $A_2$ drops to some fraction of it (e.g. half the maximum value). With two-armed spirals or elongated tidal tails present in the disc, that is not always possible as these features keep the value of $A_2$ high. One way around that would be to use the radius where $A_2$ peaks as the bar length measure. That, however,  represents a lower limit for the estimate and typically corresponds to half of the real bar extent \citep{peschken2019}. A more suitable method would be to take advantage of the fact that phase angle $\phi_2$ (Equation~\ref{eq:a2_p2}) remains roughly constant inside the bar region. Due to noise, the bar phase $\phi_2$ varies within the bar region and is not perfectly constant. To determine the bar region and thus the bar length, one would first need to define the allowed deviation of the phase from either the mean phase in the region or the phase $\phi_{2,\mathrm{max}}$ (the phase of the amplitude peak $A_{2,\mathrm{max}}$). There is no strict criterion for the allowed deviation $\Delta\phi_2$, but the value should fall within some reasonable range (e.g. $10^\circ - 20^\circ$). However, the choice of a fixed value might lead to unexpected issues. If relatively low, the length of longer bars would be realistic, but shorter bars would go undetected due to higher noise on lower radii. Alternatively, higher values of $\Delta\phi_2$ would make it possible to detect shorter bars, but it would overestimate the length of longer bars.

We thus propose the choice of allowed deviation variable with radius $R$:
\begin{equation}\label{eq:dp2}
    \Delta \phi_2 (R) = \mathrm{arcsin}\bigg(\frac{R_\mathrm{cut}}{R}\bigg)
\end{equation}
\noindent where $R_\mathrm{cut}$ is arbitrarily chosen free parameter. It represents the fixed maximum allowed distance from a fixed angle at any given radius, resulting in variable allowed angular distance on different radii. For the purpose of this work, we adopted $R_\mathrm{cut} = 1.5$ kpc, as it corresponds to $\Delta\phi_2\simeq14.5^\circ$ on $R=R_\mathrm{D}$. We outline our procedure and discuss it in detail in the list below:
\begin{enumerate}
    \item First, we locate the local maxima $\{A_{2,\mathrm{max}}\}$ (peaks of $A_2(R)$) and their corresponding radii $\{R_\mathrm{max}\}$ in inner parts of the disc $R < 2 R_\mathrm{D}$. We run the procedure for each of those. Generally, we would only detect one peak and thus run the procedure once. However, this generalisation allows us to detect possible occurrences of double bars.
    \item We then locate $R_\mathrm{min}<R_\mathrm{max}$ where the amplitude drops to half of the maximum value, $A_2(R_\mathrm{min})\simeq A_{2,\mathrm{max}}/2$. As a measure of the bar phase, we use the mean phase in the bar region instead of the phase of amplitude peak. Thus we calculate the mean phase $\overline{\phi_2}$ in $(R_\mathrm{min},R_\mathrm{max})$ range, and deviation of $\phi_2(R_\mathrm{min})$ from it. If the real deviation is lower than the maximum allowed one defined with Equation~\ref{eq:dp2}, we adopt $R_\mathrm{min}$ as a measure of the bar width. Otherwise, we increase $R_\mathrm{min}$ in iterative steps until the condition is satisfied. This iterative part rarely happens, and it is possibly necessary only for outer bars in the aforementioned scenario of double bars. Note that the condition defined with Equation~\ref{eq:dp2} is only valid for $R \geq R_\mathrm{cut}$. In case our initially located $R_\mathrm{min}$ is lower, $R_\mathrm{min}<R_\mathrm{cut}$, we simply adopt that value as a measure of the bar width without testing against this condition.
    \item In a similar manner to $R_\mathrm{min}$, we define $r_\mathrm{B}>R_\mathrm{max}$, calculate the mean phase $\overline{\phi_2}$ in $(R_\mathrm{min},r_\mathrm{B})$ range and deviation of $\phi_2(r_\mathrm{B})$ from it. We increase $r_\mathrm{B}$ in iterative steps as long as the real deviation is lower than the maximum allowed one, and adopt the highest value of $r_\mathrm{B}$ (where the condition is still satisfied) as the bar length.
    \item We implement additional, final condition for the ellipticity of the bar $\epsilon = (r_\mathrm{B}-R_\mathrm{min})/r_\mathrm{B}>0.3$, which is inspired by a similar condition used in a different bar detection method based on ellipse fitting \citep[e.g.][]{jogee2004,lee2019}, while the actual value of $0.3$ is chosen ad hoc within reasonable limits (commonly used values being $0.25$, $0.4$). Finally, if the adopted bar width $R_\mathrm{min}$ and length $r_\mathrm{B}$ satisfy this condition we consider the bar detected.
\end{enumerate}
Our procedure, while solving many of the issues we face in this work, is overly complex. It might be redundant for simple cases where usual methods would suffice (e.g. in case of stable and steady evolution of bars in the absence of spiral arms).

\subsection{RESULTS AND DISCUSSION}
\begin{figure}
\includegraphics[width=0.95\columnwidth,keepaspectratio=true]{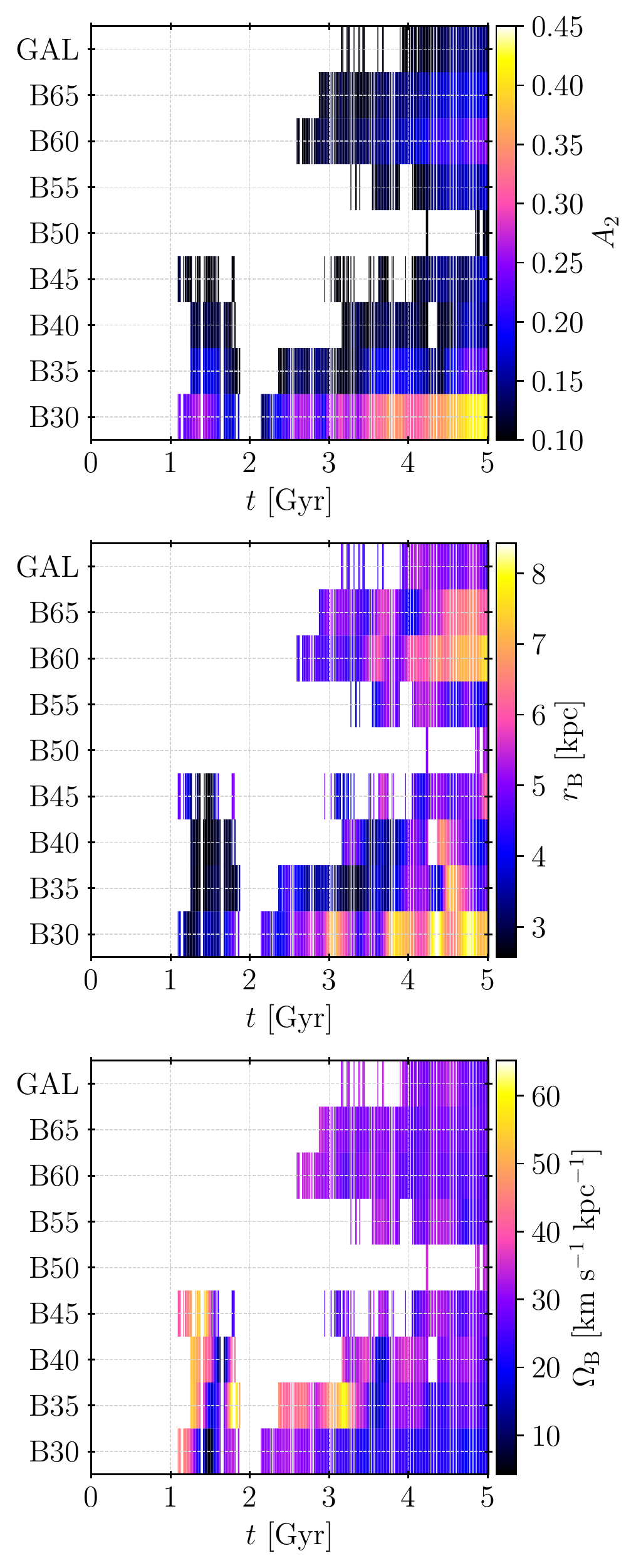}
\caption{Evolution of bar parameters in flyby simulations and isolated galaxy model (denoted as GAL), top to bottom: bar strength $A_2$, length $r_\mathrm{B}$ and pattern frequency $\Omega_\mathrm{B}$.}
\label{barmap}
\end{figure}
The evolution of bar parameters is shown in Figure~\ref{barmap}, where $A_2$ is the bar strength, $r_\mathrm{B}$ its length and $\Omega_\mathrm{B}$ its pattern frequency calculated as a temporal change of the bar phase $\overline{\phi_2}$. Our galaxy model, denoted as GAL, is unstable, with bar starting to appear around $\sim 3.15$ Gyr but only growing steadily after $t = 3.92$ Gyr. Its strength continuously grows from $A_2 = 0.1$ to $A_2 = 0.157$ at the end, which means that it remains very weak. The length of that bar does not change drastically, starting with $r_\mathrm{B} = 4.4$ kpc and varying around $\sim 4.9$ kpc during its evolution. There are also small variations in pattern frequency around the constant value of $\Omega_\mathrm{B} \simeq 30$ km s$^{-1}$ kpc$^{-1}$.

As a direct consequence of a flyby, a short bar with the length $r_\mathrm{B}\sim 3$ kpc forms almost immediately after the encounter is over (at $t = 1.09$ Gyr in simulations B30 and B45, and at $t = 1.25$ Gyr in B35 an B40) if the strength of interaction is sufficiently large, $S\geq 0.076$. Its strength $A_2$ at formation time anti-correlates with impact parameter (and thus directly correlates with strength of interaction), ranging from $A_2 = 0.24$ in simulation B30 to $A_2 = 0.12$ in B45. Highly variable pattern frequency $\Omega_\mathrm{B}$ of these bars is likely caused by the way we calculate this parameter. Specifically, defining the bar phase as a mean phase $\overline{\phi_2}$ within the bar region is somewhat unreliable due to the higher noise on small radii. This bar is short-lived as it dissolves after $<1$ Gyr, but it quickly rebuilds in stronger, closer flybys with $S\geq 0.129$ and continues to grow and evolve. By the end of simulation, at $t = 5$ Gyr, it reaches $A_2 = 0.42$ and $r_\mathrm{B} \simeq 7.7$ kpc in B30, and $A_2 = 0.26$ and $r_\mathrm{B} \simeq 5.3$ kpc in B35, both with comparable pattern frequency $\Omega_\mathrm{B} \simeq 22$ km s$^{-1}$ kpc$^{-1}$, lower (i.e. slower) than the isolated counterpart.

While \citet{lokas2018} investigated equal-mass flybys (i.e. with mass ratio $q=1$), four different interaction strengths explored $S = \{0.02,0.07,0.15,0.26\}$ make it suitable, to an extent, for comparison with our study. In their simulations, bars formed on prograde orbits in three strongest interactions, the bar strength correlates with interaction strength, and there was a brief period of bar dissolution. Our results mostly support those findings except for the weaker interactions. For example, in our simulations B40 and B45, it is unclear whether the later bar formation is caused by initially formed but dissolved bar rebuilding itself or accelerated pre-exisitng bar instability. In particular, in B40 the bar re-emerges again at $t=3.16$ Gyr and grows (with a brief dissolution around $t = 4.25$ Gyr) until the end, reaching the strength $A_2 = 0.15$ and length $r_\mathrm{B} \simeq 5.3$ kpc. On the other hand, in B45, the bar starts to appear again around $t = 2.94$ Gyr but only grows steadily after $t = 4.04$ Gyr, reaching the final strength  $A_2 = 0.18$ and length $r_\mathrm{B} \simeq 6.3$ kpc.

Contrary to our results, \citet{lang2014} found that the bar forms only in the secondary galaxy in flybys with a mass ratio $q=0.1$. They do not provide information on the strength of interaction. However, based on the data given, we can estimate that the interaction with $q=1$ is almost four times stronger than with $q=0.1$. With the strength of interaction accounted for, their results (and weak bar-like feature they identified with ellipse fits in the primary galaxy in $q=0.1$ flyby) become understandable. For example, despite differences between our simulations in initial models and interaction setup, we can compare their results with our simulation B40. And indeed, the bar that initially forms in that simulation is very weak and short-lived. Later bar formation and evolution in B40 is most likely a result of pre-existing bar instability in our galaxy model.

Interestingly, this pre-existing bar instability is suppressed in simulation B50, having only brief periods when the weak bar (with the length $r_\mathrm{B}\simeq5$ kpc and strength $A_2=0.1$) appears. As it starts happening consistently after $t = 4.84$ Gyr, it is possible that bar formation happening in isolation is significantly delayed in this simulation. In B55, the bar formation and evolution is mildly affected by the flyby interaction. The bar starts to appear more consistently earlier than in isolation, around $t=3.54$ Gyr, ending its evolution slightly stronger ($A_2 = 0.18$), shorter ($r_\mathrm{B}\simeq 4.7$ kpc) and slower ($\Omega_\mathrm{B} \simeq 27$ km s$^{-1}$ kpc$^{-1}$) than its isolated counterpart. Furthermore, bar formation is significantly accelerated in both B60 and B65, starting at $t =2.59$ Gyr and $t = 2.87$ Gyr, respectively. In both simulations final bars are stronger, longer and slightly slower than in isolation. The values are $A_2 = 0.24$, $r_\mathrm{B}\simeq 7.4$ kpc and $\Omega_\mathrm{B} \simeq 24$ km s$^{-1}$ kpc$^{-1}$ in simulation B60, and $A_2 = 0.18$, $r_\mathrm{B}\simeq 7.5$ kpc and $\Omega_\mathrm{B} \simeq 26$ km s$^{-1}$ kpc$^{-1}$ in B65.

Hence, farther and weaker flybys have all the possible effects on the pre-existing bar instability: its suppression or delay, acceleration, and even almost no effect at all. All of these effects are known to emerge from tidal interactions \citep{gerin1990,lokas2016,pettitt2018,zana2018,peschken2019}. However, there is no clear and uniform correlation with the impact parameter (or, alternatively, interaction strength). We argue that the consequences of galaxy flybys can be of primary and secondary nature. Primary, direct ones happen early and show a uniform correlation (or anti-correlation) with the impact parameter (e.g. spiral arms, angular momentum gain and its redistribution). Those can then, through various internal processes, lead to secondary, indirect effects and, evidently, vastly different outcomes. As such, these secondary effects do not necessarily depend directly on the impact parameter or interaction strength, as a combination of various internal processes can appear chaotic.

\subsection{INTERESTING CASE OF MULTIPLE STRUCTURES}
\begin{figure}
\includegraphics[width=0.85\columnwidth,keepaspectratio=true]{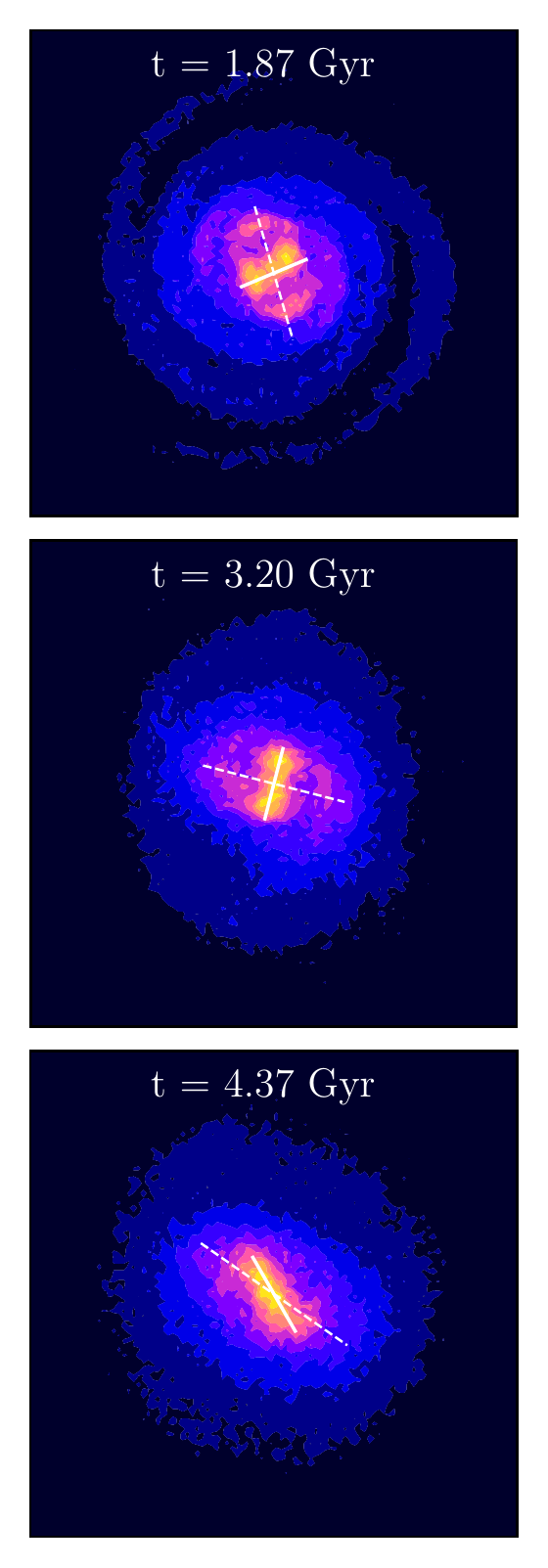}
\caption{Examples of face-on disc projections (of an inner disc where $R<20$ kpc) in simulation B35 at three different times (top to bottom) specified on each picture. White solid lines correspond to the major axis of the main bar while white dashed lines represent the major axis of the secondary structure.}
\label{b35pics}
\end{figure}
\begin{figure}
\includegraphics[width=0.95\columnwidth,keepaspectratio=true]{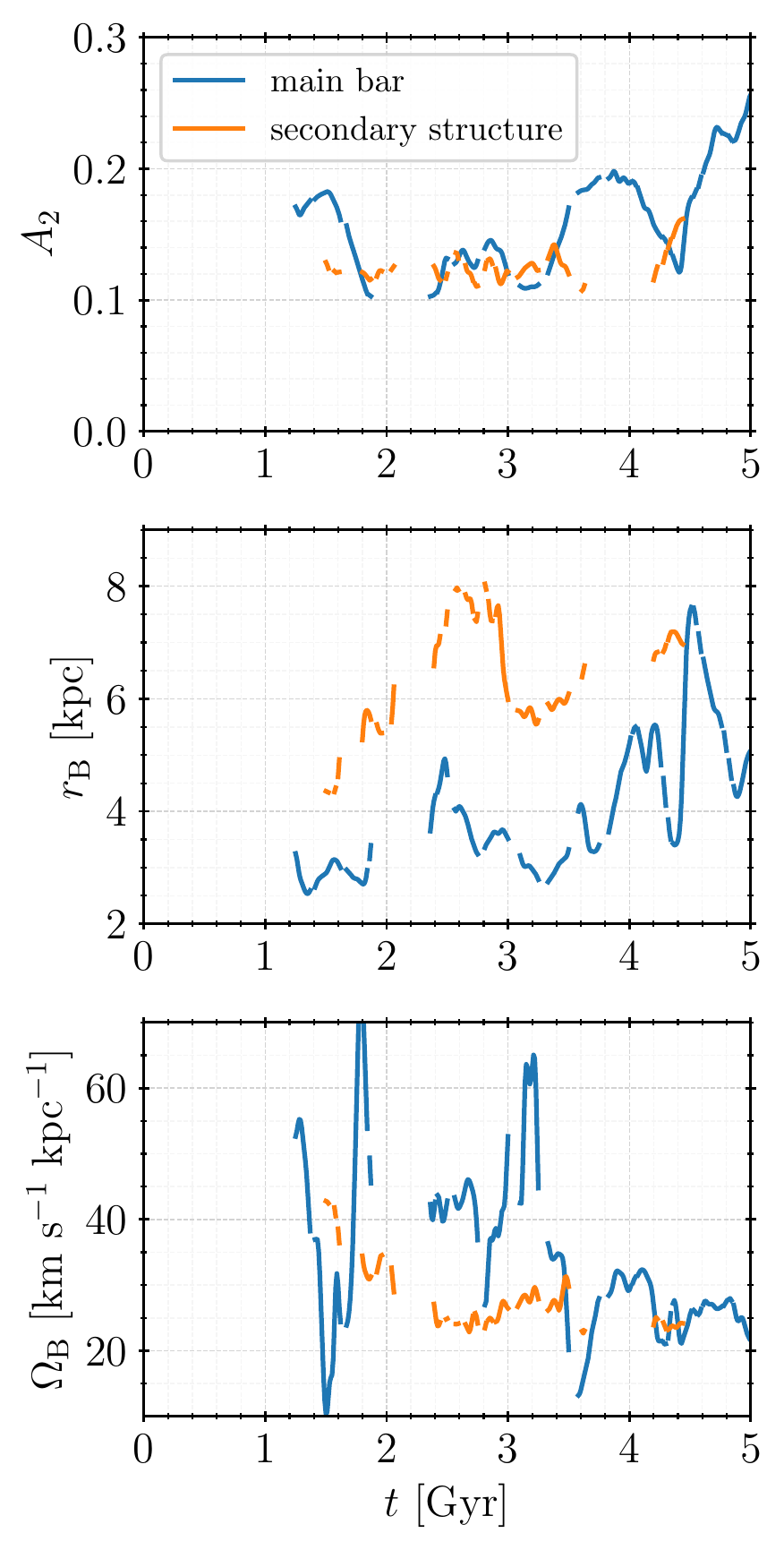}
\caption{Evolution of bar parameters (as shown in Figure~\ref{barmap}) in simulation B35. The blue lines represent the main bar, while orange ones represent outer, secondary structure.}
\label{b35both}
\end{figure}
Simultaneous multiple structures we identified as bars (in some simulations) can manifest visually as either a double bar (also known as a bar within the bar) or a ring-like feature around the main bar. Such occurrences are present in simulations B30 and B40 only for short periods, but simulation B35 is particularly interesting as these two structures coexist for an exceptionally long time. Thus, we will showcase this phenomenon in B35 here, in an illustrative manner as detailed analysis is out of the scope of this paper.

While observational studies indicate that around one-third of barred galaxies are double-barred \citep{erwin2002,laine2002,erwin2004}, their formation is still puzzling. Numerous formation mechanisms have been proposed, which could be classified into two cases: inner bars form after gas inflow through the outer bar \citep{friedli1993,heller2001,shlosman2002,englmaier2004} or they can be formed dynamically from inner discs \citep{debattista2007,shen2009,du2015}. The latter case does not restrict the order of the formation time of two bars: inner bars can form before the outer ones.

In our case, the origin of the outer structure identified as a bar is that of a rapidly evolving spiral wrapping around the slowly-evolving early-formed bar. More specifically, as a bar forms early on in closer, stronger flybys and the extent of induced spiral structure is larger, starting on lower radii, inner parts of the evolving spirals wrap around the bar. This initially leads to a ring-like feature around the main bar but can evolve into a double bar. Illustrative examples of face-on disc projections of an inner disc (where $R<20$ kpc) in simulation B35 at three different times when such structures are present are shown in Figure~\ref{b35pics}. White solid lines represent the major axis of the main bar, and white dashed lines the major axis of the secondary structure. Spiral origin is the most evident in the early stages, at $t = 1.87$ Gyr, as the secondary structure has an almost ring-like or arc-like shape. Later on, as these two structures evolve, at $t = 3.20$ Gyr, the spiral origin is less evident, but two major axes are still perpendicular. Finally, in later stages at $t = 4.37$ Gyr, this combined structure becomes a distinct double-barred feature.

The evolution of bar parameters (as shown in Figure~\ref{barmap}) for these two structures in simulation B35 is shown in Figure~\ref{b35both}. Interestingly, the strengths $A_2$ of the two structures are comparable, and as expected, the secondary structure has systematically higher lengths $r_\mathrm{B}$. Despite being highly variable, the pattern frequency $\Omega_\mathrm{B}$ of the main bar is higher than that of the secondary structure, which allows them to coexist and co-evolve for such a long time as separate structures. Moreover, this result seems to be in good agreement with \citet{debattista2007} who noticed that, for dynamically formed double bars, inner bars rotate faster. At $t = 3.63$ Gyr, these two structures sync, merging into a single bar, which results in abrupt growth of the main bar (which gets stronger, longer and starts rotating slower). Multiple structures reappear at $t = 4.2$ Gyr only to sync and merge again at $t = 4.44$ Gyr. After this final sync, the main bar grows in strength but gets shorter without any significant change in its pattern frequency.

Despite \citet{lokas2018} not reporting about this interesting feature, such occurrences may be present as well in their simulations (albeit for shorter periods), especially considering Figure 5 in their paper, which shows the evolution maps of $A_2(t, R)$ in all simulations. However, without the evolution map of the phase $\phi_2(t, R)$, we cannot claim this is certainly the case.

\subsection{THE OBSERVABILITY OF THESE BARS}
\begin{figure}
\includegraphics[width=0.95\columnwidth,keepaspectratio=true]{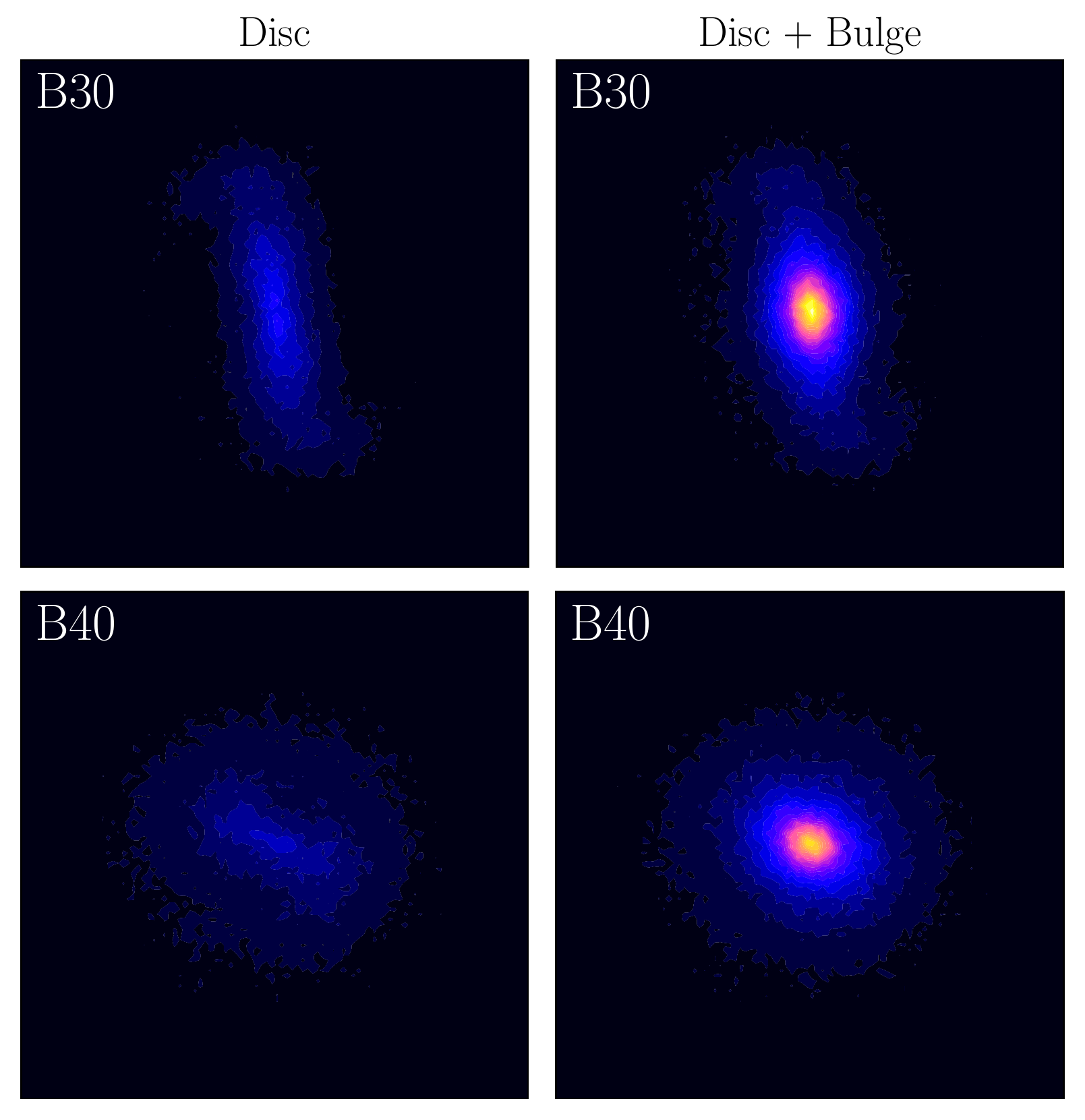}
\caption{Surface densities of disc particles only (left panels) and disc with bulge particles included (right panels) for two different simulations: B30 (upper panel) and B40 (lower panels), at $t=5$ Gyr, in the inner disc region where $R<15$ kpc. All pictures are normalised to the absolute maximum density that appears $\Sigma = 2.03 \cdot 10^8$ M$_\odot$ kpc$^{-3}$.}
\label{obspic}
\end{figure}
During our analysis and results, one could easily notice we considered disc particles only. This is mainly because, in the presence of a massive bulge, subtle changes in the disc would not be detectable. However, that raises the question: are these short or weak bars even observable? For short bars (e.g. when $r_\mathrm{B}\leq 4$ kpc), the answer should be obvious as those bars are fully embedded into the radius which encloses $70\%$ of the bulge's mass. For longer but weaker bars, the answer is not as clear. As an example, in Figure~\ref{obspic} we show surface densities, side by side, of the disc particles only and the disc with bulge particles included for two different simulations, B30 and B40, at $t = 5$ Gyr, in the inner disc region where $R<15$ kpc. All pictures are normalised to the absolute maximum density that appears $\Sigma = 2.03 \cdot 10^8$ M$_\odot$ kpc$^{-3}$. We chose these simulations in particular, as they illustrate this the best: the bar in B30 is long and strong enough, while the one in B40 is somewhat long but rather weak. Stronger bars (e.g. when $A_2>0.2$) can still be detected in the presence of a massive bulge, but their shape becomes more ocular. On the contrary weaker bars, despite their length, are not observable. Although not easily detectable and observable, these bars should still be able to contribute to the angular momentum transfer and, in the presence of gas, gas inflow to the galaxy centre. Inner bars, like the one we showcased in the previous example of multiple structures, are particularly efficient in transporting gas and can even trigger nuclear activity \citep{shlosman1989,shlosman1990}.

The ability to easily separate the subsystems of the galaxy, disc and bulge, is evidently an attractive perk of numerical simulations. In practice, while observing the real galaxies, this is not the case, and the presence of a bulge can contaminate the analysis and make reliable bar detection even more challenging. This introduces additional problems for research on the role of bars in AGN fuelling. An extensive number of studies were performed on this topic, with conflicting results \citep[e.g.][]{laine2002,coelho+11,oh+12,alonso13,alonso14,cheung+15,galloway+15,goulding+17,Silva-Lima2022}. These conflicting results might be explained partially by the differences in detection methods for both the nuclear activity and bars. For example, \citet{lee2019}, while studying bright galaxies from SDSS/DR7, found that the fraction of bars is dependent on the detection method (visual inspection, ellipse-fitting, Fourier method). Based on our results, we argue that there might be an additional problem, related to the sample selection since some bars are not observable and could not be detected with any direct, visual method.
\begin{figure*}
\includegraphics[width=0.9\textwidth,keepaspectratio=true]{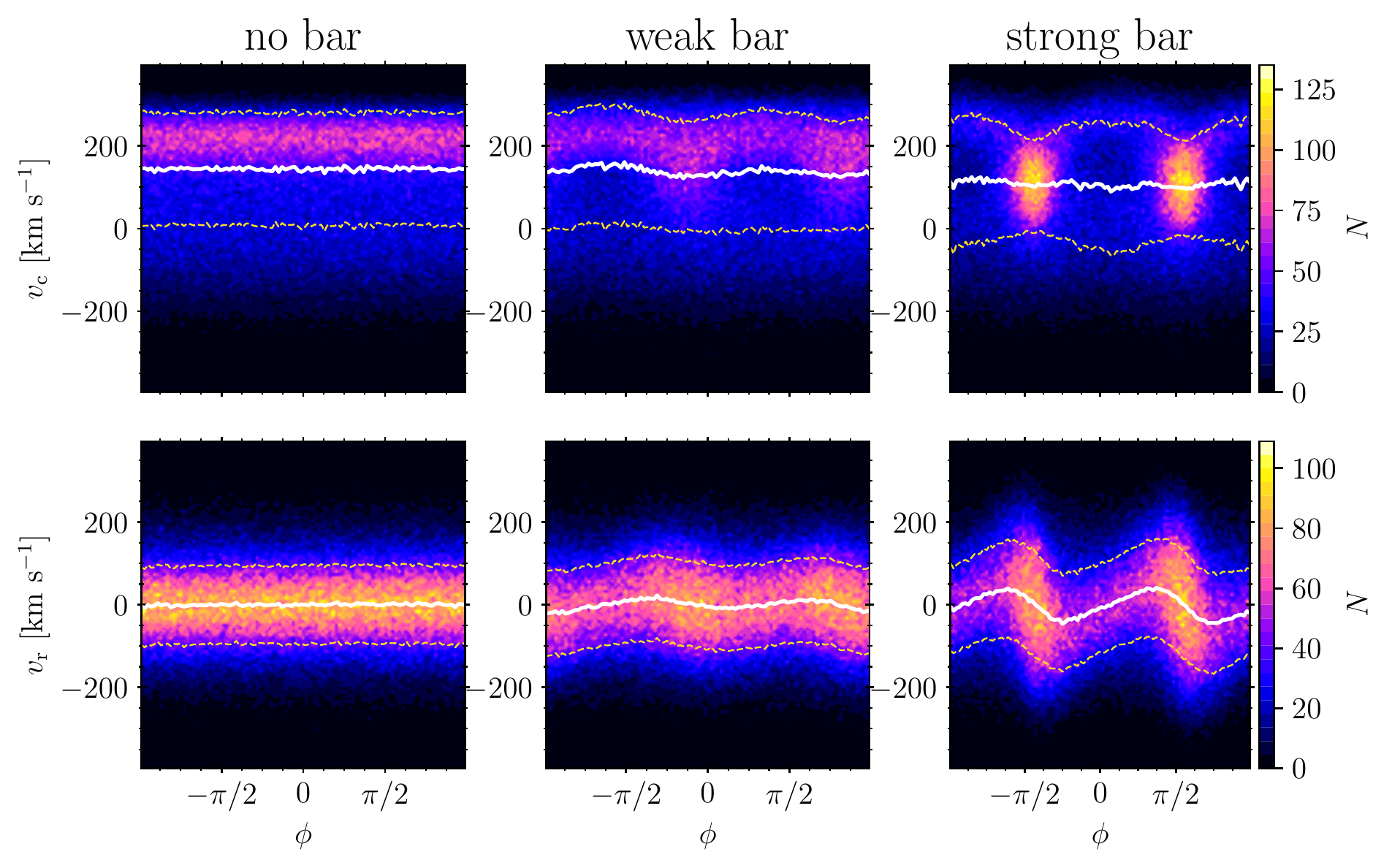}
\caption{Circular velocity $v_c$ (\emph{upper panels}) and radial velocity $v_r$ (\emph{lower panels}) distributions across azimuth $\phi$ of all stellar particles (disc + bulge) within half mass radius of the stellar component ($\sim 8$ kpc) for three scenarios (\emph{left to right}): no bar, weak bar and strong bar. Solid white lines represent median values and dashed yellow ones dispersion.}
\label{kinmaps}
\end{figure*}

Still, it might be possible to detect these weak bars indirectly. Analysing stellar kinematics, if such data are readily available or obtainable, would be one of the possibilities. Stellar orbits are, typically, elongated along the bar resulting in detectable non-circular (i.e. radial) motions and the velocity dispersion in central regions of barred galaxies is higher than in its non-barred counterpart \citep[e.g.][]{Kormendy1983,Bettoni+1988}. As an example, in Figure~\ref{kinmaps}, we show the distributions of all stellar particles (disc + bulge) residing inside half mass radius of the stellar component along azimuth $\phi$, for both circular $v_c$ (upper panels) and radial velocity $v_r$ (lower panels). We showcase three different scenarios (left to right): no bar, weak bar and strong bar. Solid white lines represent median values, while dashed yellow ones represent dispersion. Expectedly, a non-barred scenario has constant median values for both velocity types along the azimuth, with radial velocity $v_r$  median value around zero, and constant dispersion. The wavy pattern of radial velocity in the case of a strong bar is evident and also expected \citep[e.g.][]{Bettoni&Galletta1997,athanassoula2002}. However, a weak bar case also shows deviations from the symmetry, making it possible to detect a weak bar as such. Although not directly measurable in observational studies, there are techniques for separating these velocity components \citep[e.g.][]{Maciejewski2012,XookSuut2021arXiv} which are not limited to stars and can be successfully implemented to explore the gas content of galaxies, its non-circular motions and flows \citep{Lopez-Coba2022}.

A major drawback of our study is, evidently, the lack of gas. We already mentioned that the presence of gas could affect our results, making spiral patterns more pronounced and longer-lived due to enhanced star formation. There are, however, multiple ways it can affect our results in the context of bars. In gas-rich galaxies, during the early stages of bar evolution and its growth, the influx of gas to the centre (through the bar region) is expected \citep{Berentzen+2007}. This naturally leads to enhanced star formation, especially in the central region \citep{Diaz-Garcia+2020,Lin+2020}, which can make bars stronger and eases their detection. The effect is more pronounced in stronger and fast-growing bars, which leads to efficient gas depletion with shorter timescales \citep{Geron+2021}. Thus, in the later stages of bar evolution, especially in the case of strong bars, we can expect quenching of star formation and lower gas fractions in the central region compared to the non-barred counterparts. It is necessary to explore the co-evolution of spiral arms and bars, which is particularly important in the presence of gas. Namely, strong spiral arms can also drive gas inflow to the centre as recently demonstrated by \cite{Yu+2022}, which can lead to additional enhancement of the central star formation and the build-up of pseudo-bulges or faster bar growth.

\section{THE EFFECT ON SPHERICAL COMPONENTS}\label{sec:bulgehalo}
\begin{figure*}
\includegraphics[width=0.99\textwidth,keepaspectratio=true]{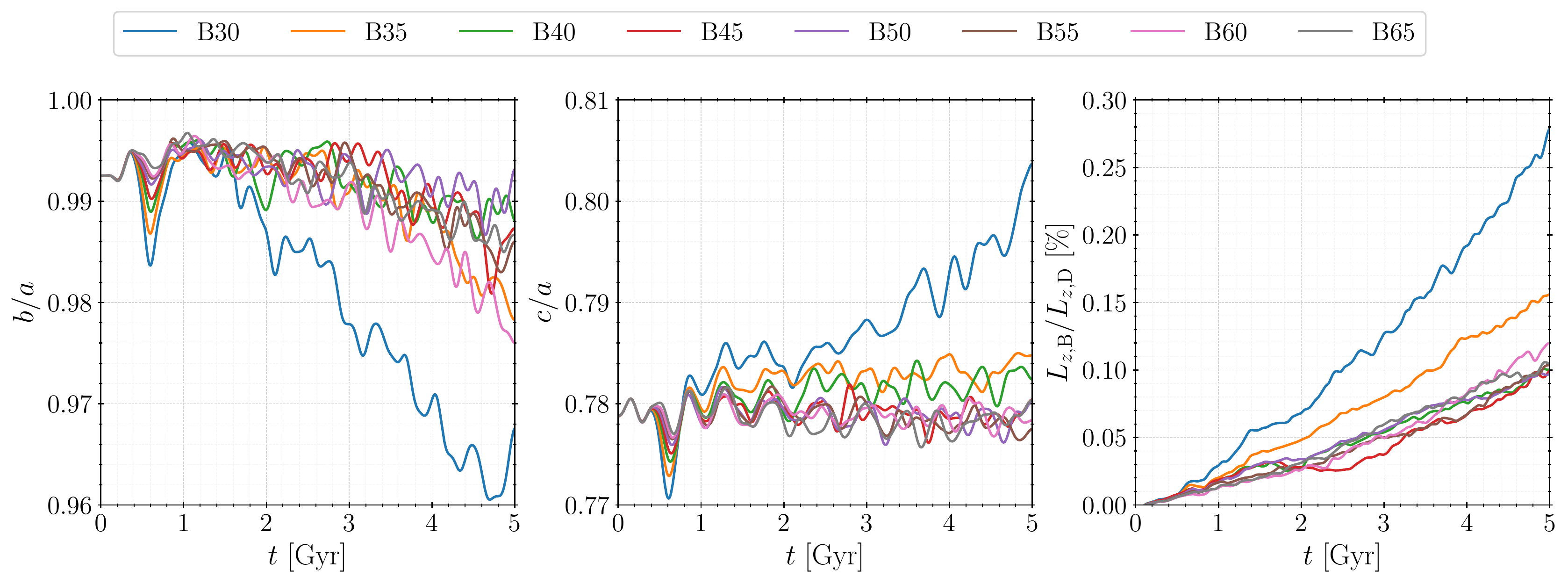}
\caption{Evolution of the bulge, left to right: axis ratio $b/a$, $c/a$ and $z$-component of the angular momentum relative to the $z$-component of the angular momentum of the disc $L_{z,\mathrm{B}}/L_{z,\mathrm{D}}$, given in percentages. Different simulations are represented with different line colours, as indicated by the legend.}
\label{bulgeevolucija}
\end{figure*}
\begin{figure}
\includegraphics[width=0.95\columnwidth,keepaspectratio=true]{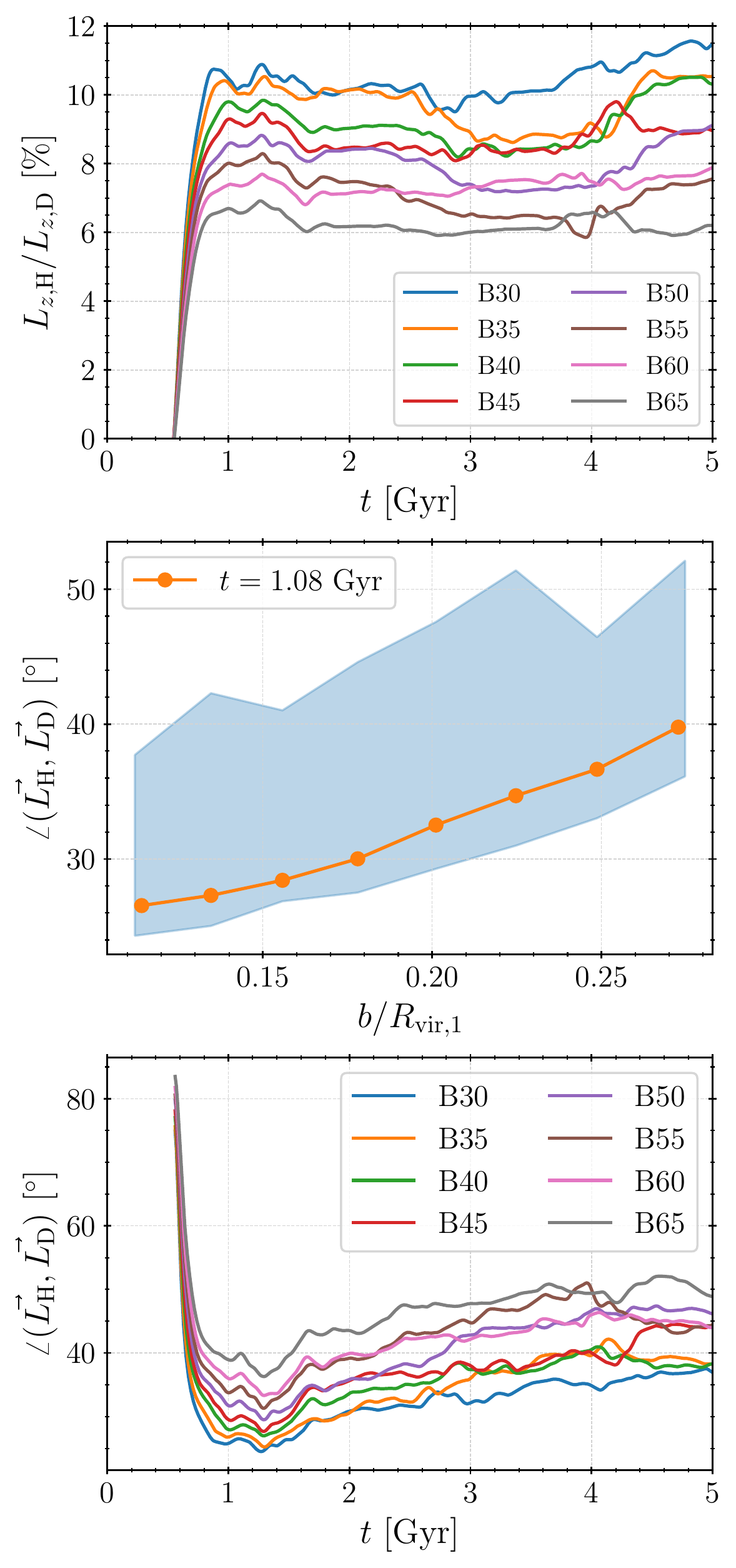}
\caption{\emph{Upper panel:} Evolution of the $z$-component of the angular momentum of the dark matter halo relative to the $z$-component of the angular momentum of the disc $L_{z,\mathrm{H}}/L_{z,\mathrm{D}}$, given in percentages. Different simulations are represented with different line colours. \emph{Middle panel:} Offset angle between the angular momenta of the dark matter halo and the disc after the encounter as a function of the relative impact parameter. Orange circles represent values in different simulations at the end of the encounter $t = 1.08$ Gyr, and the blue-coloured area further variations of this offset after the encounter. \emph{Lower panel:} Evolution of the offset angle between the angular momenta of the dark matter halo and the disc after the pericentre. Different simulations are represented with different line colours.}
\label{haloevolucija}
\end{figure}
Classical bulges are quite resilient to flybys interactions, as already demonstrated by \citet{kumar2021}. However, for completion's sake, we analysed the evolution of the bulge's shape and $z$-component of the angular momentum vector relative to the $z$-component of the angular momentum of the disc $L_{z,\mathrm{B}}/L_{z,\mathrm{D}}$. We show the evolution of these parameters in Figure~\ref{bulgeevolucija}, where $a$, $b$, and $c$ are the lengths of the longest, intermediate, and shortest axis, respectively, of the bulge's mass distribution. While the bulge remains largely unchanged in most flyby simulations, simulation B30 stands out. The bulge's shape is mildly affected, and the final spin-up corresponds to $\sim0.28\%$ of the disc's angular momentum. Both of these effects are not directly caused by the flyby interaction, but the rapidly-evolving strong and long bar that forms in that simulation and the bar-bulge interaction \citep[e.g.][]{kataria2018,kataria2019}. To further emphasise this, we remind the reader that the shortest axis of the mass distribution $c$ is not necessarily aligned with the angular momentum vector of the bulge. However, the actual offset angle between them is only a few degrees ($2^\circ-4^\circ$), and the angular momentum of the bulge is almost perfectly aligned with the disc's angular momentum vector. Moreover, the spin-up of the bulge is perfectly gradual throughout the simulations, with no signs of sharp and abrupt angular momentum gain during the interaction. This suggests that the bulk of the bulge's newly gained spin is transferred from the disc and that the intruder is too far away (even at its pericentre) to affect the bulge directly.

Aside from angular momentum, a non-negligible amount of mass is transferred between the disc (or, specifically, the bar) and the bulge, which affects the bulge's shape. Using a tool for the structural decomposition of galaxies based on stellar kinematics \citep[e.g. \texttt{MORDOR} algorithm,][]{MORDOR2022} could help us estimate the amount of transferred mass since the majority of particles associated with the bar would be flagged as a pseudo-bulge component. A strong bar in simulation B30, which accounts for roughly $20\%$ of the total disc mass (or around $30\%$ of the inner disc, where $R<15$ kpc), captures up to $6\%$ of the bulge particles. Based on the trends seen in Figure~\ref{bulgeevolucija}, specifically the evolution of the bulge's axis ratios, it is likely that the amount of the bulge's mass captured by the bar is going to increase until a distinct pseudo-bulge is formed.

Due to its extended nature, the initially non-rotating dark matter halo is the first to experience flyby effects. The intruder galaxy transfers its orbital angular momentum to the spin angular momentum of the dark matter halo during the encounter. At first, these two angular momenta are aligned due to the so-called spin-orbit alignment \citep{Moon2021, An2021}. The initially gained angular momentum of the dark matter halo is almost perpendicular to the angular momentum of the disc (Figure~\ref{haloevolucija}, lower panel). However, this does not last long, and the halo spin adjusts to more or less align with the disc spin before the encounter is over. We show the evolution of the $z$-component of the angular momentum of the dark matter halo relative to the $z$-component of the angular momentum of the disc $L_{z,\mathrm{H}}/L_{z,\mathrm{D}}$ in Figure~\ref{haloevolucija} on the upper panel. The actual amount of the angular momentum gain of the dark matter halo is somewhat higher (due to the offset angle between the vectors of angular momenta of these two subsystems), and it varies between $14.3\%$ in the simulation with the closest flyby (B30) and $9.9\%$ in the one with the farthest (B65). Thus, there is a clear anti-correlation (of halo spin-up) with the impact parameter.

The offset angle between the angular momenta of the dark matter halo and the disc after the encounter $\angle (\vec{L_\mathrm{H}},\vec{L_\mathrm{D}})$, as a function of the relative impact parameter, is shown on the middle and its evolution on the lower panel in Figure~\ref{haloevolucija}. Orange circles (middle panel) represent values in different simulations at the end of the encounter, at $t = 1.08$ Gyr, and the blue-coloured area further variations of this offset after the encounter. It is clear that the halo spin never perfectly aligns with the disc spin and that there is an almost linear correlation between this offset and impact parameter. Variations in this offset angle are mostly comparable between simulations, with an increasing trend in all of them. However, note that we considered here the whole dark matter halo. The inner parts are usually prone to the much more frequent short-term changes of spin orientation \citep{bett2012}. As these changes can affect the morphological evolution of the galaxy, this effect could also contribute to the unusual and non-uniform effects farther, weaker flybys have on the pre-existing bar instability. By the end of simulations, the offset angle between the angular momenta of the dark matter halo and the disc is correlated with the impact parameter, ranging from about $35^\circ$ in B30 to about $50^\circ$ in B65 (Figure~\ref{haloevolucija}, lower panel). Different layers of the dark matter halo may have different spin orientations. For example, the inner parts, despite expected frequent short-term spin-flips, could constantly re-align with the disc spin, while the outermost parts can retain their initial perpendicular orientation, which results in an overall offset of around $40^\circ$.

\begin{figure}
\includegraphics[width=0.95\columnwidth,keepaspectratio=true]{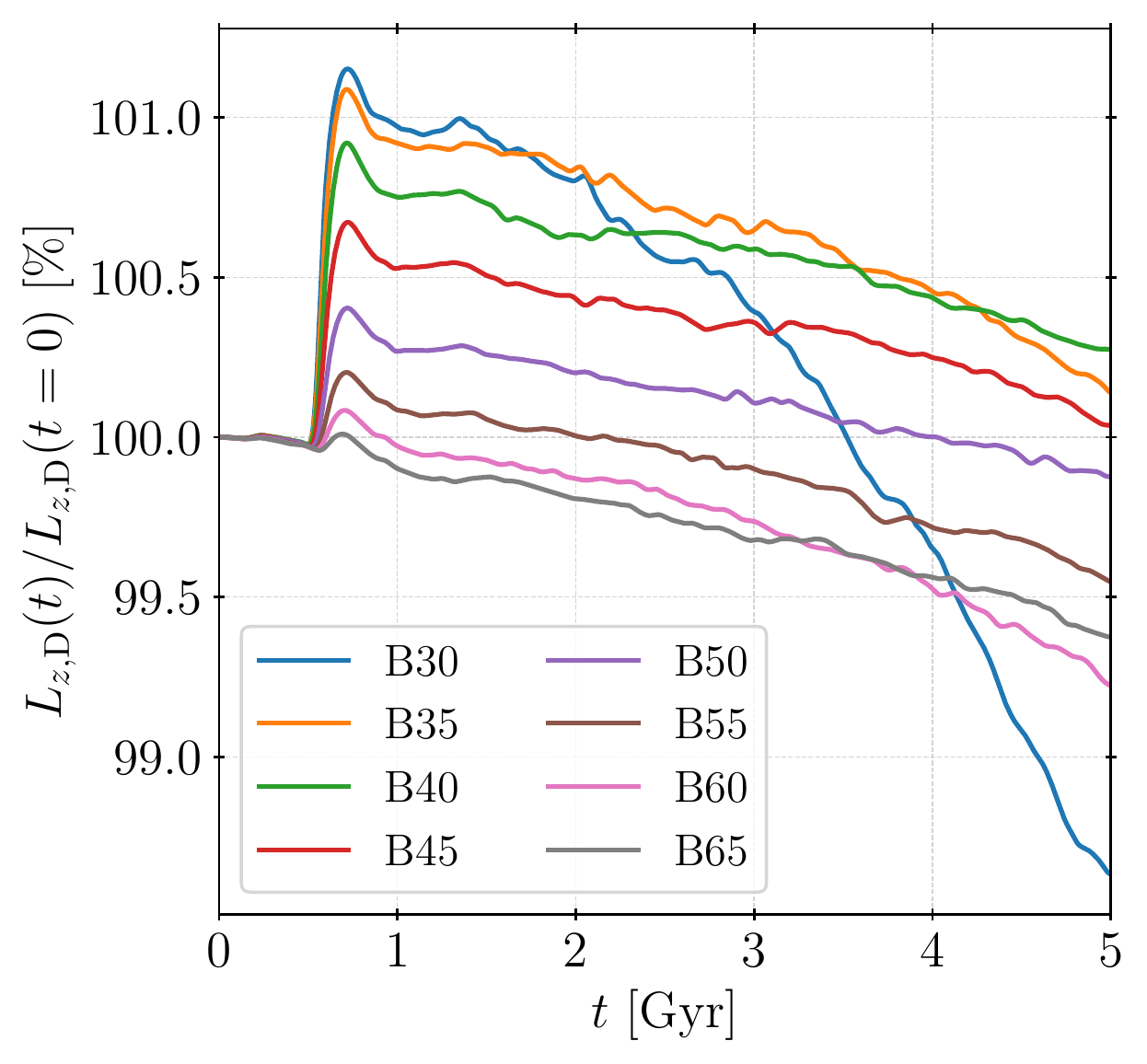}
\caption{Evolution of the $z$-component of the angular momentum of the disc relative to its initial value (at $t = 0$) $L_{z,\mathrm{D}} (t)/L_{z,\mathrm{D}}  (t=0)$, given in percentages. Different simulations are represented with different line colours.}
\label{Lzd-evol}
\end{figure}

We show the evolution of the $z$-component of the angular momentum of the disc relative to its initial value $L_{z,\mathrm{D}}(t)/L_{z,\mathrm{D}}(t=0)$ in Figure~\ref{Lzd-evol}. Note that this $z$-component of the angular momentum represents the total angular momentum of the disc since we rotate the whole system aligning the angular momentum vector of the disc with the $z$-axis. After the pericentre, the disc gains angular momentum anti-correlated with the impact parameter, up to $\sim 1.2 \%$ of its initial value in simulation B30. However, the disc does not retain its newly gained angular momentum. Instead, it transfers it to both spherical components (dark matter halo and stellar bulge), losing the angular momentum, until the end, at a constant pace in almost all simulations. Simulation B30, again, stands out. The loss of angular momentum of the disc is much steeper and, by the end of the simulation, the disc has $\sim 98\%$ of its initial angular momentum. This is due to the strong bar that forms in that simulation, which efficiently transfers angular momentum to both spherical components. However, while there is a correlation between angular momentum transfer and the impact parameter (and B30 is a drastic, special case), the change in angular momentum of the disc is about $2\%$ at most (over $5$ Gyr), making this effect relatively minor on a global scale.

\begin{figure*}
\includegraphics[width=0.9\textwidth,keepaspectratio=true]{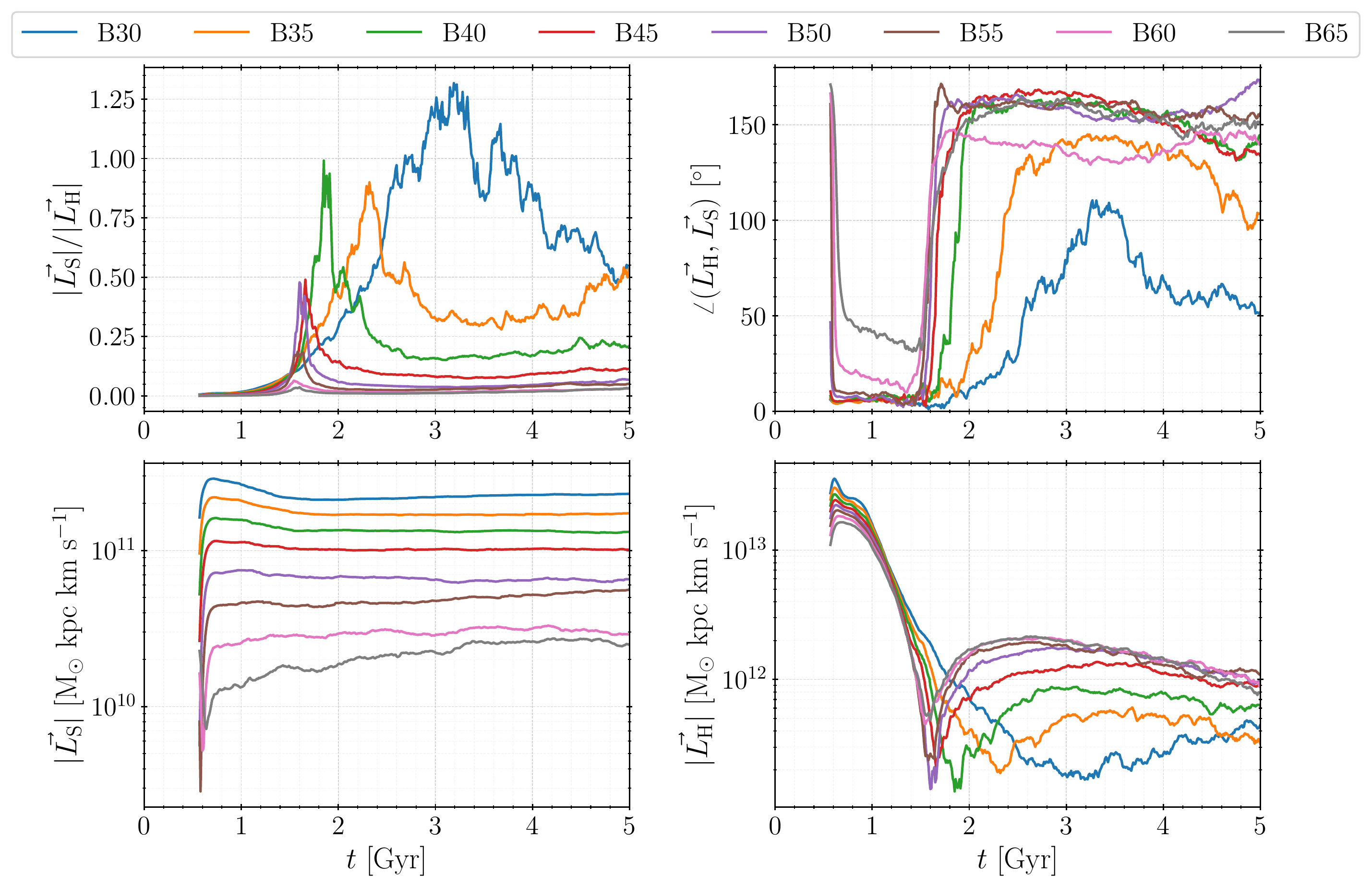}
\caption{Angular momentum evolution of the secondary galaxy, where $\vec{L_\mathrm{S}}$ is the angular momentum of the stellar component (\emph{lower left panel}) and $\vec{L_\mathrm{H}}$ that of the dark matter halo (\emph{lower right panel}). Ratio of their intensities is shown in upper left panel, and the angle between them in upper right panel.}
\label{int-angmom}
\end{figure*}
Finally, the angular momentum evolution of the secondary, intruder galaxy (stellar $\vec{L_\mathrm{S}}$ and dark matter halo ones $\vec{L_\mathrm{H}}$, ratio of their intensities and the angle between those two vectors) is shown in Figure~\ref{int-angmom} for completion's sake. The stellar component, which consists only of one spherical subsystem, gains a non-negligible amount of angular momentum after the pericentre, which remains almost constant (slightly increases during the simulation in B65) until the end. This angular momentum gain is anti-correlated with the impact parameter and hence, correlated with interaction strength. Similarly, the intruder's dark matter halo also gains angular momentum after the pericentre anti-correlated with the impact parameter. However, it is followed by the immediate and steep angular momentum loss, which coincides with the period of intense tidal stripping of the dark matter halo \citep{mitrasinovic2022}. During this period, two angular momentum vectors are almost aligned in most simulations (B30-B55). After the dark matter halo stabilises, its virial mass becomes almost constant (i.e. no mass loss), and it regains a certain amount of angular momentum. In this phase, the offset between two angular momentum vectors is high ($\angle (\vec{L_\mathrm{H}},\vec{L_\mathrm{S}}) > 140^\circ$ in most simulations with farther flybys). This suggests that the perfect observational candidates for galaxies residing in counter-rotating dark matter haloes are lower-mass galaxies in denser environments \citep[given that a frequency of flybys is higher in such environmental regimes, as demonstrated by][]{shan2019}, in the local Universe, i.e. at lower redshifts \citep{sinha2012,shan2019}.

\section{IMPLICATIONS AND OVERALL DISCUSSION}
\label{sec:impdis}
We demonstrated that frequent, typical flybys with lower mass ratios significantly affect the evolution of galaxies, producing various observed effects. Our results are similar to those of previous studies which focused on flybys with higher mass ratios \citep[e.g.][]{lang2014,lokas2018}, when we compare interactions based on their strengths. Since the effects on the primary galaxy are essentially the same for a given strength of the interaction, there are no observable signatures that would help us differentiate low-mass ratio flybys from high-mass ratio ones, in a post-flyby stage.

Ongoing close interaction is typically observed as a close galaxy pair. Detecting these visually interacting pairs may be biased towards interactions of higher mass ratios and massive galaxies in dense environments \citep[e.g.][]{Lotz+2010,Blumenthal+2020}. Moreover, it is not always possible to correctly classify the interaction type of a galaxy pair based on the outcome (into a merger or non-merger). To do so, one would need precise measurements of their physical separation and relative velocity. This process is significantly easier and more direct in numerical simulations, although misclassification can still happen. For example, \cite{shan2019} mentioned that as high as $\sim 20\%$ of galaxy pairs that appear gravitationally unbound and would, thus, be classified as flyby if no other condition is applied, end up merging. Thus, for ongoing close interaction, even if it can be successfully observed as a close pair, it is not always possible to assess whether the interaction is a merger or a close flyby.

The mass ratio of interacting galaxies is one of the most important parameters describing galaxy mergers. Given that mergers of different mass ratios produce severely different effects, it is understandable that the most common way to classify them (into major or minor) is based on this parameter. However, this is not the case for flybys or any non-merger interaction in general. The mass ratio plays a significant role in identifying the ongoing interaction but, when it comes to the produced effects, it only plays a part given that interaction strength depends on a few other parameters as well (impact parameter, as explored in this work, and interaction timescale). It is, thus, more viable to classify non-mergers based on the interaction strength \citep[ideally with][parameter, since it takes into consideration the interaction timescale]{elmegreen1991}, especially in theoretical works where the procedure is practical. Classifications based on a single parameter (for example, major/minor based on the mass ratio or distant/close based on the impact parameter) neglect a crucial role of the interaction strength and may give results that appear counter-intuitive.

Cosmological simulations clearly show that the fraction of flybys becomes significant, compared to the fraction of mergers, at lower redshifts $z\leq2$ and that flybys can even outnumber mergers, especially for massive, Milky Way-like primary galaxies \citep{sinha2012}. Aside from the higher frequency of flybys at lower redshifts, \citet{shan2019} also reported that there is a strong correlation between environmental density and the number of flybys: in the densest environments (such as big clusters of galaxies), and for Milky way-like primary galaxies, flybys outnumber mergers by order of magnitude. Non-merger interactions are, generally, more frequent than mergers in dense environments such as groups or clusters of galaxies \citep{gnedin2003}, where some galaxies can experience multiple and, on average, around $30\%$ of galaxies experience at least one flyby per orbit \citep{Knebe2004PASA}. Field galaxies can also experience flybys, despite less frequently. This reinforces the idea that flybys are as relevant as ever, for the structural evolution of galaxies, in the local, present-day Universe.

The most obvious and observable signature of galaxy flybys is the formation of distinct, tidally induced features, such as two-armed spirals and bars. Distinct two-armed spirals form in all flybys, which appear as grand design spirals in most strong enough flybys (e.g. $S\geq 0.05$), while long-lived bars form only in stronger flybys ($S\geq 0.129$). Note that our work is focused only on the effects of a single flyby interaction and, as such, it is mostly applicable to the field galaxies. In dense environments, galaxies are prone to constant external perturbations: from multiple flybys, satellite galaxies and even global environmental effects. Examining luminous face-on spiral galaxies from the COSMOS survey, \citet{Sheth+2008} found that the fraction of barred galaxies is highly dependent on the redshift, sharply increasing from $\sim 20\%$ at $z\sim 0.84$ to $\sim 65\%$ in the local Universe, which was confirmed in cosmological simulations \citep[e.g.][]{Cavanagh+2022,MORDOR2022}. Such a high difference cannot be explained by the flybys alone. Flybys that are strong enough to produce long-lived bars account for only $\sim 20\%$ of total flybys for massive galaxies \citep{sinha2015}. However, weaker flybys can still induce mild bar instabilities, which can lead to bar formation if combined with other effects in dense environments. Thus, despite not being the sole contributor to the high population of barred galaxies in the local Universe, close galaxy flybys still play a significant role in the build-up of bars in the local galaxies.

The environment can also affect the flyby-induced two-armed spirals. For example, a galaxy with two-armed, grand-design spirals, which initially formed due to flyby interaction, residing in a denser environment is less likely to lose those spiral arms due to their decay. Instead, its spiral structure is going to get revitalised by environmental effects \citep{semczuk2017}, further contributing to the higher fraction of grand design spiral galaxies in dense environments. On the contrary, the flyby-induced spiral structure of the isolated field galaxy is going to gradually decay, since there is a lack of external perturbations and, in the later stages, it might appear flocculent. This is in line with the observational reports \citep[e.g.][]{Elmegreen1982} that find flocculent spirals are the most common among isolated galaxies without companions, while grand design spirals are predominantly found in denser environments. Moreover, the tidal origin of distinct two-armed grand design spirals is well known \citep{kendall2015,Hart+2018,Sellwood2022ARA&A}. There are, however, puzzling examples of isolated grand design spiral galaxies that challenge this formation theory \citep[e.g.][]{Elmegreen1982,kendall2011,kendall2015}. In most of our simulations, distinct two-armed spirals are visible long after the intruder has left the proximity of the primary galaxy (i.e. the distance between them can exceed $1$ Mpc). This suggests that we should not easily dismiss the tidal origin of these puzzling cases of isolated grand design spirals, as they are the best observational candidates for galaxies that experienced close flyby in their recent history.

\section{SUMMARY AND CONCLUSIONS}
\label{sec:sum}
\indent Equal-mass flybys, those with mass ratios $q=1$ or close, sparked more interest, especially regarding bars and spirals. However, one has to keep in mind that such flybys are extremely rare and almost exclusively distant \citep{sinha2012,sinha2015}. Frequent, typical flybys have mass ratios $q=0.1$ or lower with secondary galaxy penetrating deep into the primary. This can usually result in comparable strengths of interaction $S$ between the two classes of flybys, and lead to essentially the same effects. To demonstrate this, we performed a series of $N$-body simulations of typical flybys with a 10:1 mass ratio and variable impact parameter $b/R_{\mathrm{vir},1}$ ranging from $0.114$ to $0.272$ of the virial radius of the primary galaxy, which corresponds to the strength of interaction $S$ ranging from $S=0.177$ in the closest flyby to $S=0.034$ in the farthest. We evolved the system for 5 Gyr and studied the evolution of morphological features long after the encounter. Our main results and conclusions are as follows.
\begin{enumerate}
    \item The disc thickens as a result of a decrease in scale length $R_\mathrm{D}$ and an increase in scale height $z_\mathrm{D}$. The effect is more pronounced in closer, stronger flybys. Which of these two changes contributes more to the thickening highly depends on a morphological structure that forms in the disc: spiral arms drive the change of the scale length, while bars significantly affect the scale height.
    \item Two-armed spirals form in all simulations before the encounter is even over. Their radii of origin correlate with the impact parameter, and their maximum strengths are well approximated with an inverted S-curve (a good alternative is a cubic regression). The impact parameter does not affect the shapes and lifetimes of well-resolved spirals. Lifetime periods of well-resolved spiral arms are $T_\mathrm{LF}\sim 2$ Gyr, and the spirals are detectable long after the intruder has left the system. However, lifetime periods of the spiral remnants or weakly-resolved spirals are longer as the impact parameter increases. 
    \item As a direct consequence of flyby, a short and weak bar can form after the encounter is over for impact parameters $b/R_{\mathrm{vir},1} \leq 0.178$  and corresponding interaction strengths $S\geq0.076$. Its length appears constant $r_\mathrm{B}\sim 3$ kpc, while the strength anti-correlates with the impact parameter. Such feature is short-lived ($<1$ Gyr) except for stronger interactions with $S\geq0.129$ when it continues to grow and evolve with a brief period of bar dissolution.
    \item We observed an interesting feature of multiple bar-like structures in simulation with interaction strength $S=0.129$. As inner parts of evolving spirals wrap around the early-formed bar, they form a ring-like structure which evolves into a double bar feature that survives for an exceptionally long time. For these two structures to coexist as separate for such a long time, the criteria are strict: the induced spiral arms have to start on low enough radii, and the early-formed bar has to evolve slowly but rotate rapidly.
    \item Effects on the pre-existing bar instability, that develops much later, are diverse: from an acceleration of bar formation, little to no effect, to even bar suppression. There is no uniform correlation between these effects and the impact parameter, as they are secondary effects happening later in a post-flyby stage. Primarily induced effects (e.g. spiral arms, angular momentum gain and its redistribution) heavily influence pre-existing mild bar instability leading to these secondary effects.
    \item Classical bulges are resilient to flyby interactions, but they can experience mild changes. However, this is not a direct result of the flyby but a by-product of bar-bulge interaction.
    \item Dark matter halos can significantly spin up in the amount anti-correlated with the impact parameter. There is an offset angle between the angular momentum vector of the dark matter halo and that of a disc, and it correlates linearly with the impact parameter.
    \item Intruder galaxy also gains a non-negligible amount of angular momentum during the interaction. In the later post-flyby stage, when the episode of intense tidal stripping is over, the intruder stabilises, and its dark matter halo is almost counter-rotating compared to the stellar component.
\end{enumerate}
We demonstrated that the effects of typical flybys with lower mass ratios could be just as significant as those of equal-mass flybys. These low mass ratio flybys started gaining more attention recently due to their ability to produce dark matter-deficient or ultra-diffuse galaxies out of secondary galaxies. One has to keep in mind that primary galaxies can also get affected and altered. This is particularly important in the local, present-day Universe when the frequency of flybys is at its highest. Moreover, close galaxy flybys can explain the puzzling observational examples of isolated two-armed grand design spiral galaxies, given that the tidal formation mechanism is preferred for such structures.

\begin{acknowledgement}
We thank the anonymous reviewer for insightful comments and suggestions that improved the quality of our paper. This work was supported by the Ministry of Science, Technological Development and Innovation of the Republic of Serbia (MSTDIRS) through the contract no. 451-03-47/2023-01/200002 made with Astronomical Observatory of Belgrade. AM acknowledges also Faculty of Mathematics, University of Belgrade, contract number 451-03-47/2023-01/200104. The python packages \texttt{MATPLOTLIB} \citep{Hunter2007}, \texttt{NUMPY} \citep{Harris2020}, \texttt{SCIPY} \citep{Virtanen2020}, \texttt{PANDAS} \citep{McKinney2010}, and \texttt{PYNBODY} \citep{pynbody} were all used in parts of this analysis.
\end{acknowledgement}


\bibliography{example}

\appendix

\end{document}